*Chapter 5*

# NON-HAMILTONIAN NATURE OF NUCLEON DYNAMICS IN AN EFFECTIVE FIELD THEORY

## *Renat Kh.Gainutdinov[1] and Aigul A.Mutygullina*

Department of Physics, Kazan State University, 18 Kremlevskaya St, Kazan 420008, Russia

## ABSTRACT

Problems connected with non-Hamiltonian nature of low energy nucleon dynamics in the effective field theory (EFT) of nuclear forces is investigated by using the formalism of the generalized quantum dynamics (GQD) developed in [J. Phys. A , 5657 (1999)]. This formalism is based on a generalized dynamical equation derived as the most general equation of motion consistent with the current concepts of quantum physics. By using the example of the EFT of nuclear forces, we demonstrate that a theory, which, being formulated in terms of Hamiltonian formalism, leads to ultraviolet divergences, may manifest itself as a perfectly consistent theory free from infinities, if it is considered from the more general point of view provided by the GQD. We show that non-Hamiltonian character of nucleon dynamics gives rise to some new problems connected with discontinuity of the evolution operator. This discontinuity results in the fact that the Hilbert space of nucleon states cannot be realized in the standard way. A space which allows one to realize, in a natural way, the Hilbert space of nucleon states is investigated. The structure of this space reflects the existence of the high energy degrees of freedom which affect on low energy nucleon dynamics.

## INTRODUCTION

The assumption that the dynamics of a quantum system is governed by the Schrödinger equation is a dynamical postulate of the Hamiltonian formalism. Quantum mechanics based on this postulate provides an excellent description of atomic phenomena. For this reason it seems natural to expect that low energy nucleon dynamics should also be governed by the

---

[1] E-mail: Renat.Gainutdinov@ksu.ru

Schrödinger equation with a nucleon-nucleon ($NN$) potential. However, such a $NN$ potential, which could play the same role as the Coulomb potential in atomic physics, has not constructed yet. Nowadays there exist "realistic" $NN$ potentials which successfully describe two-nucleon scattering data to high precision, but they are not sufficiently strong to reproduce many-nucleon data. The main reasons for this is that, despite the quark and gluon degrees of freedom are not observable in low energy regime, they have significant effects on low energy nucleon dynamics. A first attempt to construct a bridge between QCD describing the quark-gluon dynamics and low energy nuclear physics was made by Weinberg [1]. He suggested to derive a $NN$ potential in time-ordered chiral perturbation theory (ChPT). Following the pioneering work of Weinberg, the effective field theory (EFT) approach which is an invaluable tool for computing physical quantities in the theories with disparate energy scales has become very popular in nuclear physics (for a review, see Ref.[2]). To describe low energy processes involving nucleons and pions, in the EFT of nuclear forces all operators consistent with the symmetries of QCD are included in an effective Lagrangian. This theory is organized as an expansion in $Q/\Lambda$, where $Q$ is a momentum scale which characterizes the process under consideration and $\Lambda$ is the range of validity of the effective theory. A fundamental difficulty is that the effective Lagrangians yield graphs, which are divergent and give rise to singular quantum mechanical potentials. To resolve this problem one has to use some renormalization procedure, which regulates the integrals and subtracts the infinities. In this way one can successfully perform calculations of many quantities in nuclear physics. However, in this case one cannot parametrise the interactions of nucleons, by using some Lagrangian or Hamiltonian, and there are not any equations for renormalized amplitudes in the subtractive EFT of nuclear forces. The effective Lagrangians of the theory are only of formal importance: They include only bare parameters, while the renormalized ones are introduced on the final stage of calculations.

The above problem of EFT's is the same that arises in any quantum field theory with ultraviolet (UV) divergences: Regularization and renormalization allow one to render the physical predictions finite, however, it is impossible to construct a renormalized Hamiltonian acting on the Fock space, i.e., after renormalization the dynamics of the theory is not governed by the Schrödinger equation. This equation is local in time, and the interaction Hamiltonian describes an instantaneous interaction. On the other hand, locality is the main cause of UV divergences in quantum field theory (QFT), and hence regularization and renormalization may be considered as some ways of nonlocalization of the theory. The fact that nonlocalization could resolve the UV divergence problem was realized long ago [3]. However, in quantum field theory, if we spread the interaction in space, we spread it in time as well, with consequent loss of causality or unitarity. Note, in this connection, that one of the basic ingredient of quantum field theory is quantum machanics. More precisely, quantum field theory is based on the same quantum mechanics that was invented by Schrödinger, Heisenberg, Pauli, Born, and others in 1925-26 [4]. In this formulation of quantum mechanics it is assumed that dynamics of a quantum system is governed by the Schrödinger equation, i.e., is generated by an instantaneous interaction. Thus the UV divergences problem in quantum field theory proceeds from its quantum-mechanical ingredient, and, in order to resolve this problem one has to find a way of solving the evolution problem in the case when the dynamics of a system is generated by a nonlocal-in-time interaction. This is possible only if the Schrödinger equation is not the basic dynamical equation of quantum theory that must be satisfied in any case. Meanwhile, in the Feynman formulation of quantum mechanics [5,6] dynamics of a quantum system is described without resorting to the Schrödinger equation. Feynman's theory starts with analysis of the phenomenon of quantum interference which leads directly to the principle of the superposition of probability amplitudes. According to this principle, the probability amplitude of an event which can occur in several different ways is a sum of probability amplitudes for each of these ways [5]. The Feynman formulation also contains, as its essential idea, the concept of a probability amplitude associated with a completely specified path in space-time, and it is postulated that this probability amplitude is an exponential whose phase is the classical action (in units of $\hbar$) for the path in question. Using this postulate together with the above principle of the superposition leads to Feynman's sum-over-path formalism. It should be emphasized that the second postulate is not so fundamental as the principle of the superposition and is introduced to make the theory equivalent to the canonical quantum mechanics that rests on the Schrödinger equation. The price for this benefit is the same too: the definition of Feynman's integrals in terms of a time interval derived into infinitesimal interval pieces which is needed in this case makes the Feynman theory local in time, despite the fact that this theory is basically global in character. In Ref.[7] it has been shown that there is another way of using the Feynman superposition principle that allows one to preserve the global character of the theory: instead of the second postulate, together with this principle one can use the first principles of the canonical formalism that manifest the probabilistic character of quantum mechanics and establish the connection between the vectors and operators and states of a quantum system and observables. In this way an equation of motion that is more general than the Schrödinger equation has been derived [7]. Being equivalent to the Schrödinger equation in the case of instantaneous interactions, this generalized dynamical equation permits the generalization to the case where the dynamics of a quantum system is generated by a nonlocal-in-time interaction. It has been shown [7] that a generalized quantum dynamics (GQD) developed in this way may be an important tool for solving various dynamical problems in quantum physics [8,9].

The main result obtained in Ref.[7] is that in general the dynamics of a quantum system should be governed by the generalized dynamical equation, and there are no physical reasons to restrict ourselves to the case of local interactions where this equation is equivalent to the Schrödinger equation. In other words, the situation where the dynamics of a closed quantum system is generated by a nonlocal-in-time interaction is possible in principle. In Ref.[8] it has been shown that this possibility is realized in the case of low energy nucleon dynamics, and in leading order of the EFT approach this dynamics is governed by the generalized dynamical equation with a nonlocal-in-time interaction operator. Moreover, this dynamics is just the same as in the case of the model developed in Refs.[7,10] as a test model demonstrating the possibility of going beyond Hamiltonian dynamics provided by the GQD.

Thus even in the nonrelativistic limit we cannot restrict ourselves to the Hamiltonian dynamics. In fact, as it follows from the Weinberg analysis of diagrams in ChPT, QCD leads to the low energy theory in which the Schrödinger equation makes no sense without regularization and renormalization. This means that the low energy predictions of QCD are inconsistent with the ordinary quantum mechanics based on the assumption that the dynamics of a quantum system is governed by the Schrödinger equation. At the same time, after renormalization the two-nucleon $T$ matrix obtained by summing the time-ordered diagrams in ChPT satisfies the generalized dynamical equation with a nonlocal-in-time interaction operator. Thus the above divergence problem is the cost of trying to describe the dynamics of low energy theory produced by QCD in terms of Hamiltonian formalism while it is really non-Hamiltonian. In other words, the low energy theory of the $NN$ interaction consistent with the symmetries of QCD is inconsistent with Hamiltonian dynamics. At the same time, if we consider the problem from the GQD point of view, we see that the low energy theory produced by QCD is free from UV divergences, and the generalized dynamical equation allows one to construct the $T$ matrix and the evolution operator without regularization and renormalization. The above provide a better understanding of what is quantum mechanics as a basic ingredient of quantum field theory, and show that the generalized dynamical equation which allows one to spread interactions in time opens new possibilities for solving the problem of the UV divergences. However, the non-Hamiltonian character of the dynamics generated by a nonlocal-in-time interaction gives rise to some new problems connected with discontinuity of the evolution operator.

As it follows from Stone's theorem the assumption that the dynamics of a quantum system is governed by the Schrödinger equation is equivalent to the assumption that the evolution operator describing this dynamics is strongly continuous. In Ref.[7] it has been shown, the requirement of the strong or week continuity of the evolution operator is not on the physical grounds. It is enough to require that the matrix elements of this operator for physically realizable states are continuous [11]. For example, there are normalized states with infinite energy in the Hilbert space [11]. Such states are not physically realizable. It is reasonable to consider the states for which $\left\| H_0 \mid \psi \right\rangle \right\| < \infty$, where $H_0$ is the free Hamiltonian, as physically realizable. Thus the matrix elements of the evolution operator for all vectors $\mid \psi \rangle \in D(H_0)$, where $D(H_0)$ being domain of $H_0$, must be continuous. At the same time, if $D(H_0)$ is dense in the Hilbert space, then, as it can be proved, the matrix elements of the evolution operator for all states of this space are continuous, i.e., the evolution operator is weakly and hence strongly continuous. Let us consider the Hilbert space $H$ of a system of two nonrelativistic spinless particles. It is usually assumed that this space can be realized as the space $L^2(M)$ of square integrable functions $\psi(\mathbf{p})$, where $M$ denotes the momentum space, and $\mathbf{p}$ being the relative momentum of the particles. Since $D(H_0)$ is dense in $L^2(M)$, from the above requirements of the physical continuity it follows that the evolution operator defined on this space is strongly continuous. This means that the Hilbert space, describing states of two nucleon system in the EFT of nuclear forces cannot be realized as the space $L^2(M)$.

In this chapter we discuss the physical and mathematical aspects of the problem of discontinuous of the evolution operator and related problem of realization of the Hilbert space of states in a theory whose dynamics is generated by a nonlocal-in-time interaction. We will investigate this problem by using the example of the EFT of nuclear forces. It will be shown that the discontinuity of the evolution operator results in the fact that the Hilbert space of nucleon states cannot be realized in the standard way. We will show how to construct a space which allows one to realize, in a natural way, the Hilbert space of nucleon states. The structure of this space reflects the existence of the quark and gluon degrees of freedom which affect on low energy nucleon dynamics. By using an illustrative example, we will demonstrate that a theory, which, being formulated in terms of the Hamiltonain formalism, leads to UV divergences, may manifest itself as a perfectly consistent theory free from infinities, if it is considered from the more general point of view provided by the GQD.

The chapter is organized as follows. We begin by discussing correspondence between continuity of the evolution and the character of the dynamics of a quantum system. We then review the principal features of the GQD. Low energy predictions of ChPT are investigated in the second part of the chapter. We show that the GQD allows one to formulate the EFT of nuclear forces as a completely consistent theory free from UV divergences. Finally we discuss the problem of realization of the Hilbert space of nucleon states which arises as a consequence of discontinuity of the evolution operator.

## DISCONTINUITY OF THE EVOLUTION OPERATOR AND THE POSSIBILITY OF GOING BEYOND HAMILTONIAN DYNAMICS.

Let us consider the problem of discontinuity of the evolution operator that arise in a theory whose dynamics is generated by a nonlocal-in-time interaction. As is well known, the basic concept of the canonical formalism of quantum mechanics is that the theory can be formulated in terms of vectors of a Hilbert space and operators acting on this space. This formalism rests on the following postulates, which establish the connection between these mathematical object and observables and prescribe how to compute the probability of an event:

(i) The physical state of a system is represented by a vector (properly by a ray) of a Hilbert space.

(ii) An observable A is represented by a Hermitian hypermaximal operator $\alpha$. The eigenvalues $a_r$ of $\alpha$ give the possible values of A. An eigenvector $\left| \varphi_r^{(s)} \right\rangle$ corresponding to the eigenvalue $a_r$ represents a state in which A has the value $a_r$. If the system is in the state $\left| \psi \right\rangle$, the probability $P_r$ of finding the value $a_r$ for A, when a measurement is performed, is given by

$$P_r = \left\langle \psi \mid P_{V_r} \mid \psi \right\rangle = \sum_s |\left\langle \varphi_r^{(s)} \mid \psi \right\rangle|^2,$$

where $P_{V_r}$ is the projection operator on the eigenmanifold $V_r$ corresponding to $a_r$, and the sum $\Sigma_s$ is taken over a complete orthonormal set $\left| \varphi_r^{(s)} \right\rangle$ (s=1,2,...) of $V_r$. The state of the system immediately after the observation is described by the vector $P_{V_r} \left| \psi \right\rangle$.

These assumptions are the main assumptions on which quantum theory is founded. They describe the properties of a quantum system at fixed time $t_0$ only. In the canonical formalism the time evolution of the system is described by the evolution equation

$$\left| \psi(t) \right\rangle = U(t, t_0) \left| \psi(t_0) \right\rangle, \tag{1}$$

where $\left|\psi(t)\right\rangle$ is a state vector, and $U(t,t_0)$ is the unitary (for an isolated system) evolution operator

$$U^+(t,t_0)U(t,t_0) = U(t,t_0)U^+(t,t_0) = 1, \tag{2}$$

satisfying the composition law

$$U(t,t')U(t',t_0) = U(t,t_0), \quad U(t_0,t_0) = 1. \tag{3}$$

The evolution operator $U(t,t_0)$ has a natural decomposition

$$U(t,t_0) = 1 + iR(t,t_0), \tag{4}$$

where the unit operator represents the no-interaction part; its matrix elements are delta functions which make the final momenta the same as the initial momenta. Here we use the interaction picture where the state vectors of non-interacting particles do not vary with the time. This operator relates to the evolution operator $U_s(t_2,t_1)$ in the Schrödinger picture as $U_s(t_2,t_1) \equiv \exp(-iH_0 t_2)U(t_2,t_1)\exp(iH_0 t_1)$. Here and below we use the units in which $\hbar = c = 1$, and $H_0$ is the free Hamiltonian. In the case of an isolated system, the operator $U_s(t_2,t_1)$ depends on the difference $(t_2 - t_1)$ only, so that the operators $V(t) \equiv U_s(t,0)$ constitute an one-parameter group of unitary operators, with the group property

$$V(t_1 + t_2) = V(t_1)V(t_2), \quad V(0) = 1.$$

The above are only the general properties of the evolution operator which directly follow from the evolution equation (1) and the conservation of probability in an isolated system. For describing the time evolution of a quantum system, we need some dynamical principle that could give rise to an equation of motion. In the Hamiltonian formalism the assumption that the evolution operator satisfies the Schrödinger equation

$$i\frac{dU(t,t_0)}{dt} = H_I(t)U(t,t_0) \tag{5}$$

is used as such a dynamical principle. Here $H_I(t)$ is an interaction Hamiltonian in the interaction picture. At the same time, this dynamical postulate can be formulated in another form. In fact, if the evolution operator is assumed to be strongly continuous, i.e., if

$$\lim_{t_2 \to t_1} \left\| V(t_2)\left|\psi\right\rangle - V(t_1)\left|\psi\right\rangle \right\| = 0, \tag{6}$$

then from Stone's theorem it follows that this one-parameter group has a self-adjoint infinitesimal generator $H$ :

$$V(t) = \exp(-iHt), \quad i\frac{d}{dt}V(t) = HV(t).$$

From this it follows that the assumption that the evolution operator is strongly continuous can be used as the dynamical postulate of the Hamiltonian formalism. Moreover, it is essential to use the assumption that this operator is weekly continuous, i.e.,

$$\langle \psi_2 \mid V(t_2) \mid \psi_1 \rangle \xrightarrow[t_2 \to t_1]{} \langle \psi_2 \mid V(t_1) \mid \psi_1 \rangle \qquad (7)$$

for any $\mid \psi_1 \rangle$ and $\mid \psi_2 \rangle$ belonging to the Hilbert space, since, according to Stone's theorem, the week continuity of the evolution operator implies its strong continuity. The advantage of such a formulation of the dynamical principle of the Hamiltonian formalism consists in the fact that in this case one may extract what is not necessary on physical grounds and need to be postulated. From the physical point of view, condition (7) must not be satisfied for all vectors belonging to the Hilbert space of physical states: It is enough to require that this condition is satisfied for any physically realizable states $\mid \psi_1 \rangle$ and $\mid \psi_2 \rangle$ [11]. Note, in this connection, that there are normalized vectors in the Hilbert space that represent the states for which the energy of a system is infinite. Such states cannot be considered as physically realizable [11], and hence the corresponding matrix elements of the evolution operator need not be continuous. Indeed, from the point of the states with infinite energy $\mid \psi_{inf} \rangle$ any time interval $\delta t$ is infinite, and hence the corresponding matrix elements of the evolution operator $U(\delta t, 0)$ must be independent of $\delta t$, i.e., must be constant

$$\langle \psi'_{inf} \mid U(\delta t, 0) \mid \psi_{inf} \rangle = \langle \psi'_{inf} \mid \psi_{inf} \rangle + i \langle \psi'_{inf} \mid R(\infty, 0) \mid \psi_{inf} \rangle.$$

This means that, for the evolution operator to be weekly and hence strongly continuous, $\langle \psi'_{inf} \mid R(\infty, 0) \mid \psi_{inf} \rangle$ must be zero. In this case the states with infinite energy do not play any role in the description of the dynamics of a quantum system. Such a situation takes place in the case where the effect of high energy physics on low energy dynamics is negligible. On the other hand, in such theories as the EFT of nuclear forces underlying high-energy physics may affects on low energy dynamics despite the separation of scales. In this case the evolution operator should be discontinuous, and hence the dynamics is non-Hamiltonian. Below we will demonstrate this point by using the example of the EFT of nuclear forces.

# GENERALIZED QUANTUM DYNAMICS

## The Generalized Dynamical Equation

Let us now briefly review the main features of the formalism of the GQD developed in Ref.[7]. The main idea of this formalism is that, instead of the assumption that the evolution operator is strongly continuous, i.e., that the dynamics of a system is governed by the Schrödinger equation, together with the above basic assumptions of the canonical formalism one can use the basic assumptions of the Feynman approach to quantum theory. Within Feynman's formalism [5,6] quantum theory is formulated in terms of probability amplitudes without resorting to the vectors and operators acting on a Hilbert space. In this approach the following assumption is used as the first basic postulate:

(iii) The probability of an event is the absolute square of a complex number called the probability amplitude. The joint probability amplitude of a time-ordered sequence of events is product of the separate probability amplitudes of each of these events. The probability amplitude of an event which can happen in several different ways is a sum of the probability amplitudes for each of these ways.

According to this assumption, the probability amplitude of an event which can happen in several different ways is a sum of contributions from each alternative way. In particular, the amplitude $\left\langle \psi_2 \,|\, U(t,t_0) \,|\, \psi_1 \right\rangle$, being the probability of finding the quantum system in the state $|\psi_2\rangle$ at time $t$, if at time $t_0$ it was in the state $|\psi_1\rangle$, can be represented as a sum of contributions from all alternative ways of realization of the corresponding evolution process. Dividing these alternatives in different classes, we can then analyze such a probability amplitude in different ways. For example, subprocesses with definite instants of the beginning and end of the interaction in the system can be considered as such alternatives. In this way the amplitude $\left\langle \psi_2 \,|\, U(t,t_0) \,|\, \psi_1 \right\rangle$ can be written in the form [7]

$$\left\langle \psi_2 \,|\, U(t,t_0) \,|\, \psi_1 \right\rangle =$$
$$= \left\langle \psi_2 \,|\, \psi_1 \right\rangle + \int_{t_0}^{t} dt_2 \int_{t_0}^{t_2} dt_1 \left\langle \psi_2 \,|\, \tilde{S}(t_2,t_1) \,|\, \psi_1 \right\rangle, \tag{8}$$

where $\left\langle \psi_2 \,|\, \tilde{S}(t_2,t_1) \,|\, \psi_1 \right\rangle$ is the probability amplitude that if at time $t_1$ the system was in the state $|\psi_1\rangle$, then the interaction in the system will begin at time $t_1$ and will end at time $t_2$, and at this time the system will be in the state $|\psi_2\rangle$. Here the interaction picture is used.

As it follows from the above postulate the probability amplitude $\left\langle \psi_2 \,|\, \tilde{S}(t_2,t_1) \,|\, \psi_1 \right\rangle$ can itself be represented as a the sum of amplitudes for each of the ways in which the subprocess with completely specified instants of the beginning and end of the interaction in a quantum system can happen. However, some supplementary assumptions about the history of the system are needed. In the Feynman approach it is assumed that this history can be represented

by some path in space-time. In this case the amplitude $\left\langle \psi_2 \mid \tilde{S}(t_2, t_1) \mid \psi_1 \right\rangle$ can be represented as a sum of contributions from all paths corresponding to processes in which the interaction begins at $t_1$ and ends at $t_2$. If we assume also that the contribution from a single path is an exponential whose (imaginary) phase is the classical action for this path (the second postulate of Feynman's theory) and substitute the expression obtained in this manner into Eq.(8), we arrive at Feynman's sum-over-paths formula for the transitions amplitudes. At the same time, in the formalism of the GQD the history of a quantum system is represented by the version of the time evolution of the system associated with completely specified instants of the beginning and end of the interaction in the system. Such a description of the history of a system is more general and require no supplementary postulates like the second postulate of the Feynman formalism. On the other hand, the probability amplitudes $\left\langle \psi_2 \mid \tilde{S}(t_2, t_1) \mid \psi_1 \right\rangle$, in terms of which the evolution of a system is described within the GQD, are used in the spirit of Feynman's theory: The probability amplitude of any event is represented as a sum of this amplitudes. In Ref.[7] it has been shown that the use of the operator formalism of the canonical approach allows one to derive a relation for the amplitudes $\left\langle \psi_2 \mid \tilde{S}(t_2, t_1) \mid \psi_1 \right\rangle$ which can be regarded as an equation of motion.

By using the operator formalism, we can represent the probability amplitudes $\left\langle \psi_2 \mid U(t_2, t_1) \mid \psi_1 \right\rangle$ by the matrix elements of the unitary evolution operator satisfying the composition law (3). Meanwhile, $\tilde{S}(t_2, t_1)$ whose matrix elements are $\left\langle \psi_2 \mid \tilde{S}(t_2, t_1) \mid \psi_1 \right\rangle$ may be only an operator-valued generalized function of $t_1$ and $t_2$, since only $U(t, t_0) = 1 + \int_{t_0}^{t} dt_2 \int_{t_0}^{t_2} dt_1 \tilde{S}(t_2, t_1)$ must be an operator on the Hilbert space. Nevertheless, it is convenient to call $\tilde{S}(t_2, t_1)$ an "operator" by using this word in a generalized sense. In the case of an isolated system the operator $\tilde{S}(t_2, t_1)$ can be represented in the form

$$\tilde{S}(t_2, t_1) = \exp(iH_0 t_2)\tilde{T}(t_2 - t_1)\exp(-iH_0 t_1).$$ (9)

As has been shown in Ref.[7], for the evolution operator $U(t, t_0)$ given by Eq.(8) to be unitary for any times $t_0$ and $t$, the operator $\tilde{S}(t_2, t_1)$ must satisfy the following equation (the generalized dynamical equation):

$$(t_2 - t_1)\tilde{S}(t_2, t_1) = \int_{t_1}^{t_2} dt_4 \int_{t_1}^{t_4} dt_3 (t_4 - t_3)\tilde{S}(t_2, t_4)\tilde{S}(t_3, t_1).$$ (10)

A remarkable feature of this equation is that it works as a recurrence relation, and allows one to obtain $\tilde{S}(t_2, t_1)$ for any $t_1$ and $t_2$, if it is known in an infinitesimal neighborhood of the point $t_2 = t_1$. Since the operators $\tilde{S}(t_2, t_1)$ describe the contributions to the evolution operator from the processes in which the interaction in the system begins at $t_1$ and ends at $t_2$, the above means that in order to construct the evolution operator it is sufficient to know the contributions to this operator from the processes with infinitesimal duration time of interaction. It is natural to associate these processes with the fundamental interaction in the system under study. This make it possible to use relation (10) as a dynamical equation. One needs only to specify the boundary condition determining the behavior of $\tilde{S}(t_2,t_1)$ in the limit $t_2 \to t_1$ and hence containing the dynamical information about the system. Denoting the contribution to the evolution operator from the processes associated with the fundamental interaction by $H_{int}(t_2,t_1)$, such a boundary condition can be written in the form

$$\tilde{S}(t_2,t_1) \xrightarrow[t_2 \to t_1]{} H_{int}(t_2,t_1) + o(\tau^\varepsilon), \tag{11}$$

where $\tau = t_2 - t_1$. The parameter $\varepsilon$ is determined by demanding that $H_{int}(t_2,t_1)$ must be so close to the solution of Eq.(10) in the limit $t_2 \longrightarrow t_1$ that this equation has a unique solution having the behavior (11) near the point $t_2 = t_1$. Within the GQD the operator $H_{int}(t_2,t_1)$ plays the same role as the interaction Hamiltonian in the ordinary formulation of quantum theory: It generates the dynamics of a system. Being a generalization of the interaction Hamiltonian, this operator is called the generalized interaction operator.

The operator $H_{int}(t_2,t_1)$ describes fundamental processes, starting from which, one can construct the evolution operator. In this process the system in the state $|\psi_1\rangle$ evolves freely up to some time $t$ when, as a result of the interaction, the state of the system is jumps abruptly into the state $|\psi_2\rangle$, and then the system evolves freely again. The contribution from this process into the evolution operator is of the form $\delta(t_2 - t_1)\langle\psi_2 \mid A(t_1) \mid \psi_1\rangle$, where the delta-function is needed for this contribution to be nonzero. Thus in this case the interaction operator should be of the form

$$H_{int}(t_2,t_1) = \delta(t_2 - t_1)A(t_1). \tag{12}$$

As has been shown in Ref. [7], the dynamical equation (10) with the boundary condition given by Eqs.(11) and (12) is equivalent to the Schrödinger equation with the interaction Hamiltonian $H_I(t) = \frac{i}{2}A(t)$. Thus the dynamics governed by Eq.(10) is equivalent to Hamiltonian dynamics in the case where the generalized interaction operator is of the form

$$H_{int}(t_2,t_1) = -2i\delta(t_2 - t_1)H_I(t_1). \tag{13}$$

Correspondingly the interaction operator in the Schrödinger picture

$$H_{int}^{(s)}(t_2 - t_1) = \exp(-iH_0t_2)H_{int}(t_2,t_1)\exp(iH_0t_1)$$

has the form

$$H_{int}^{(s)}(\tau) = -2i\delta(\tau)H_I,$$

where $H_I = \exp(-iH_0 t)H_I(t)\exp(iH_0 t)$. In this case the interaction generating the dynamics of a quantum system is instantaneous. On the other hand, there are no reasons to restrict ourselves to the case where the interaction operator is of the form (13). From the mathematical point of view, the boundary condition (11) with the operator $H_{int}(t_2, t_1)$ given by Eq.(13) is not only possible boundary condition for Eq.(10) which is a unique consequence of the representation (8) and the unitarity condition (2). The representation (8) in turn is a consequence of the first Feynman postulate that, as is well known, is formulated as a result of the analysis of the phenomenon of the quantum interference and hence is one of the most fundamental postulate of quantum theory. Thus Eq.(10) is a unique consequence of the first principles and can be considered as the most general dynamical equation consistent with the current concepts of quantum theory. Note, in this connection, that no new fundamental concepts and postulates are used in the formalism of the GQD. A novelty of this formalism consists in the fact that some basic postulates of the Feynman and canonical approaches to quantum theory are used in combination. This allows one to formulate the theory in terms of the operator $\tilde{S}(t_2, t_1)$. As has been shown in Ref.[7], being formulated in this way, the theory provides a more detailed description of the dynamics of a quantum system than the description directly in terms of the evolution operators, or in terms of Feynman's path amplitudes. In the case where the interaction operator is of the form (13), i.e., the interaction is instantaneous, the Schrödinger equation (5) for the evolution operator and Feynman's sum-over-paths formula follows from the representation (8) and Eq.(10). At the same time, the dynamical equation permits the generalization to the case where the interaction is nonlocal-in-time, i.e., the time durations of the interaction in the fundamental processes which determine the dynamics of a system are not zero. In this case the dynamics depends not only on the form of the operator $H_{int}^{(s)}(\tau)$ but also on its dependence upon the duration time $\tau$ of the interaction. However, as we have seen, only the behavior $H_{int}^{(s)}(\tau)$ in the limit $\tau \longrightarrow 0$ is relevant: Knowing the behavior of $H_{int}^{(s)}(\tau)$ in the infinitesimal neighborhood of the point $\tau = 0$ is sufficient to construct the evolution operator by using Eq.(10).

In order that the dynamical equation (10) with the boundary condition (11) have a unique solution, the operator $H_{int}(t_2, t_1)$ must be sufficiently close to its relevant solution. This means that this operator must satisfy the condition

$$(t_2 - t_1)H_{int}(t_2, t_1) \xrightarrow[t_2 \to t_1]{} \int_{t_1}^{t_2} dt_4 \int_{t_1}^{t_4} dt_3 (t_4 - t_3)H_{int}(t_2, t_4)H_{int}(t_3, t_1) + o(\tau^{\varepsilon+1}). \quad (14)$$

Note that the value of the parameter $\varepsilon$ depends on the form of the operator $H_{int}(t_2, t_1)$. Since $\tilde{S}(t_2, t_1)$ and $H_{int}(t_2, t_1)$ are only operator-valued distributions, the mathematical meaning of conditions (11) and (14) should be clarified. We will assume that the condition (11) means that

$$\left\langle\psi_2\left|\int_{t_0}^t dt_2\int_{t_0}^{t_2} dt_1\tilde{S}(t_2,t_1)\right|\psi_1\right\rangle\xrightarrow[t\to t_0]{}\left\langle\psi_2\left|\int_{t_0}^t dt_2\int_{t_0}^{t_2} dt_1 H_{int}(t_2,t_1)\right|\psi_1\right\rangle+o(\tau^{\varepsilon+2}),$$

for any vectors $|\psi_1\rangle$ and $|\psi_2\rangle$ of the Hilbert space. The condition (14) has to be considered in the same sense.

Note also that in general the interaction operator has the following form [8,9]:

$$H_{int}(t_2,t_1)=-2i\delta(t_2-t_1)H_I(t_1)+H_{non}(t_2,t_1),$$

where the first term on the right-hand side of this equation describes the instantaneous component of the interaction generating the dynamics of a quantum system, while the term $H_{non}(t_2,t_1)$ represents its nonlocal-in-time component.

## The Evolution Operator

If $H_{int}(t_2,t_1)$ is specified, Eq.(10) allows one to find the operator $\tilde{S}(t_2,t_1)$. Formula (8) can then be used to construct the evolution operator $U(t,t_0)$ and accordingly the state vector

$$|\psi(t)\rangle=|\psi(t_0)\rangle+\int_{t_0}^t dt_2\int_{t_0}^{t_2} dt_1\tilde{S}(t_2,t_1)\,|\psi(t_0)\rangle$$

at any time $t$. Thus Eq.(10) can be regarded as an equation of motion for states of a quantum system. By using Eqs.(8) and (9), the evolution operator can be represented in the form

$$\begin{aligned}
\langle n_2\,|\,U(t,t_0)\,|\,n_1\rangle&=\langle n_2\,|\,n_1\rangle\\
&+\frac{i}{2\pi}\int_{-\infty}^{\infty} dx\frac{\exp[-i(z-E_{n_2})t]\exp[i(z-E_{n_1})t_0]}{(z-E_{n_2})(z-E_{n_1})}\\
&\times\langle n_2\,|\,T(z)\,|\,n_1\rangle,
\end{aligned}\qquad(15)$$

where $z=x+iy$, $y>0$, and

$$\langle n_2\,|\,T(z)\,|\,n_1\rangle=i\int_0^{\infty} d\tau\exp(iz\tau)\langle n_2\,|\,\tilde{T}(\tau)\,|\,n_1\rangle.\qquad(16)$$

Here $|n\rangle$ are the eigenvectors of the free Hamiltonian $H_0$, i.e., $H_0\,|\,n\rangle=E_n\,|\,n\rangle$, and $n$ stands for the entire set of discrete and continuous variables characterizing the system in full. From Eq.(15), for the evolution operator in the Schrödinger picture, we get

$$U_s(t,0)=\frac{i}{2\pi}\int_{-\infty}^{\infty} dx\exp(-izt)G(z),\qquad(17)$$

where

$$\langle n_2 \mid G(z) \mid n_1 \rangle = \frac{\langle n_2 \mid n_1 \rangle}{z - E_{n_1}} + \frac{\langle n_2 \mid T(z) \mid n_1 \rangle}{(z - E_{n_2})(z - E_{n_1})}. \tag{18}$$

Equation (18) is the well-known expression establishing the connection between the evolution operator and the Green operator $G(z)$ and can be regarded as the definition of the operator $G(z)$.

The generalized dynamical equation (10) is equivalent to the following equation for the $T$ matrix [7]:

$$\frac{d\langle n_2 \mid T(z) \mid n_1 \rangle}{dz} = -\sum_n \frac{\langle n_2 \mid T(z) \mid n \rangle \langle n \mid T(z) \mid n_1 \rangle}{(z - E_n)^2}, \tag{19}$$

with the boundary condition

$$\langle n_2 \mid T(z) \mid n_1 \rangle \xrightarrow[|z| \to \infty]{} \langle n_2 \mid B(z) \mid n_1 \rangle + o(\mid z \mid^{-\beta}),$$

where $\beta = 1 + \varepsilon$, and

$$\langle n_2 \mid B(z) \mid n_1 \rangle = i \int_0^\infty d\tau \exp(iz\tau) \langle n_2 \mid \tilde{B}(\tau) \mid n_1 \rangle,$$

$\tilde{B}(\tau)$ being an arbitrary operator that has the following behavior in the limit $\tau \to 0$:

$$\langle n_2 \mid \tilde{B}(\tau) \mid n_1 \rangle \xrightarrow[\tau \to 0]{} \langle n_2 \mid H_{int}^{(s)}(\tau) \mid n_1 \rangle + o(\tau^\varepsilon).$$

Note in this connection that, while we define the operator $H_{int}(t_2, t_1)$ for any times $t_1$ and $t_2$, only its values for infinitesimal duration times $\tau = t_2 - t_1$ of interaction are relevant: Knowing the behavior of $H_{int}(t_2, t_1)$ in the infinitesimal neighborhood of the point $t_2 = t_1$ is sufficient to construct the evolution operator by solving Eq.(10). The two interaction operator $H_{int}(t_2, t_1)$ and $H'_{int}(t_2, t_1)$ are dynamically equivalent, provided

$$H_{int}(t_2, t_1) = H'_{int}(t_2, t_1) + o\left((t_2 - t_1)^\varepsilon\right), \quad \tau \to 0.$$

In fact these operators lead to the same solution of Eq.(10), i.e., generate the same dynamics. Correspondingly knowing the behavior of B(z) in the limit $\mid z \mid \to \infty$ is sufficient for obtaining a unique solution of Eq.(19). This behavior is uniquely determined by the behavior of $H_{int}^{(s)}(\tau)$ in the limit $\tau \to 0$. At the same time, the operator $H_{int}^{(s)}(\tau)$ need not be such that the Fourier transform $\int_0^\infty d\tau \exp(iz\tau) \langle n_2 \mid H_{int}^{(s)}(\tau) \mid n_1 \rangle$ exists, since it determine the behavior of the operator $\tilde{T}(\tau)$, for which such a Fourier transform must satisfy, only in the limit $\tau \to 0$. Of course, one can use a dynamical equivalent operator for which such a Fourier transform exists. The operators $\tilde{B}(\tau)$ is an example of such operators. However, it is not convenient to deal with such interaction operators. In fact, in this case one has to take care of the behavior of $H_{int}^{(s)}(\tau)$ not only in the limit $\tau \to 0$ but also in the limit $\tau \to \infty$. Nevertheless, formally we can construct the operator $B(z)$ for any $z$, by using the operator $\tilde{B}(\tau)$ which, being dynamically equivalent to the operator $H_{int}^{(s)}(\tau)$, satisfies the above requirement.

It should be noted that the $T$ matrix obtained by solving Eq.(19) satisfies the following equation:

$$\langle n_2 \mid T(z_1) \mid n_1 \rangle - \langle n_2 \mid T(z_2) \mid n_1 \rangle = (z_2 - z_1) \sum_n \frac{\langle n_2 \mid T(z_2) \mid n \rangle \langle n \mid T(z_1) \mid n_1 \rangle}{(z_2 - E_n)(z_1 - E_n)}. \quad (20)$$

This equation in turn is equivalent to the following equation for the Green operator

$$G(z_1) - G(z_2) = (z_2 - z_1) G(z_2) G(z_1).$$

This is the Hilbert identity, which in the Hamiltonian formalism follows from the fact that in this case the evolution operator (17) satisfies the Schrödinger equation, and hence the Green operator is of the form

$$G(z) = (z - H)^{-1},$$

where $H$ being the total Hamiltonian. Thus in the Hamiltonian formalism Eq.(20) is a consequence of the Schrödinger equation. On the other hand, in Ref.[7] this equation has been derived directly from the first principles of quantum physics, and should be valid even when the Schrödinger equation does not make a sense without renormalization. Thus Eq.(10) is more general than the Schrödinger equation which follows from this equation in the particular case where the interaction operator is of the form (13). Correspondingly, in this case the LS equation follows from Eq.(19). In fact, as is well known, in the Hamiltonian formalism the $T$ matrix has the following asymptotic behavior:

$$\langle n_2 \mid T(z) \mid n_1 \rangle \xrightarrow[|z| \to \infty]{} \langle n_2 \mid V \mid n_1 \rangle, \quad (21)$$

where $V$ being the potential. By letting $\mid z_2 \mid \to \infty$ and taking into account Eq.(21), from Eq.(20), we easily get the Lippmann-Schwinger (LS) equation

$$T(z) = V + V G_0(z) T(z),$$

where $G_0(z) = \sum_n \dfrac{|n\rangle\langle n|}{z - E_n + i0}$, provided the potential meets the ordinary requirements of quantum mechanics. Thus formally Eq.(19) with the asymptotic condition (21) is equivalent to the LS equation, provided the potential meets the ordinary requirements of quantum mechanics. Here, of cause, we have to keep in mind that in the Hamiltonian formalism the asymptotic condition (21) is derived as a consequence of the LS equation. From the GQD point of view, condition (21) must be satisfied for the interaction in the system be instantaneous.

Equation (19) can be regarded as a form of the generalized dynamical equation written in terms of the $T$ matrix. In some cases such a form of the generalized dynamical equation may be convenient for practical calculations.

## Exactly Solvable Model

The possibility of going beyond Hamiltonian provided by the GQD can been demonstrated by using the developed in Refs. [7,10] model which describes the evolution of two nonrelativistic particles whose interaction is separable. In this case the generalized interaction operator in the Schrödinger picture is of the form

$$\langle \mathbf{p}_2 | H_{int}^{(s)}(\tau) | \mathbf{p}_1 \rangle = \varphi^*(\mathbf{p}_2)\varphi(\mathbf{p}_1)f(\tau),\tag{22}$$

where $f(\tau)$ is some function of the time duration of interaction $\tau$. Here we use the center of mass system and denote the relative momentum by $\mathbf{p}$. Let the form factor $\varphi(\mathbf{p})$ have the following high momentum behavior:

$$\varphi(\mathbf{p}) \sim |\mathbf{p}|^{-\alpha}, \quad (|\mathbf{p}| \longrightarrow \infty).\tag{23}$$

As has been shown in Refs.[7], there is the one-to-one correspondence between UV (high momentum) behavior of the form factor and the character of the dynamics of a quantum system: In the case $\alpha > 1/2$ the interaction in the system is instantaneous and hence the dynamics is Hamiltonian, while in the case $\alpha \leq 1/2$ the interaction is necessarily nonlocal-in-time and, as a consequence, the dynamics is non-Hamiltonian. In order to clarify this point let us consider the solution of the dynamical equation (10) with the interaction operator of the form (22). Obviously, this solution have to be of the form

$$\langle \mathbf{p}_2 | T(z) | \mathbf{p}_1 \rangle = \varphi^*(\mathbf{p}_2)\varphi(\mathbf{p}_1)t(z).\tag{24}$$

Here the function $t(z)$ satisfies the equation

$$\frac{dt(z)}{dz} = -t^2(z)\int \frac{d^3k}{(2\pi)^3} \frac{|\varphi(\mathbf{k})|^2}{(z - E_k)^2}\tag{25}$$

with the asymptotic condition

$$t(z) \xrightarrow[|z| \to \infty]{} f_1(z) + o(|z|^{-\beta}), \tag{26}$$

where

$$f_1(z) = i \int_0^\infty d\tau \exp(iz\tau) f(\tau),$$

and $E_k = \frac{k^2}{2\mu}$, $\mu$ being the reduced mass. The solution of Eq.(25) with the initial condition $t(0) = g_0$, is

$$t(z) = g_0 \left( 1 - z g_0 \int \frac{d^3k}{(2\pi)^3} \frac{|\varphi(\mathbf{k})|^2}{(z - E_k) E_k} \right)^{-1}. \tag{27}$$

In the case $\alpha > \frac{1}{2}$, the function $t(z)$ tends to a constant as $|z| \longrightarrow \infty$:

$$t(z) \xrightarrow[|z| \to \infty]{} \lambda. \tag{28}$$

Thus in this case the function $f_1(z)$ must also tend to $\lambda$ in the limit $|z| \longrightarrow \infty$. From this it follows that the only possible form of the function $f(\tau)$ is

$$f(\tau) = -2i\lambda\delta(\tau) + f^{'}(\tau),$$

where the function $f^{'}(\tau)$ has no such a singularity at the point $\tau = 0$ as the delta function. In this case the generalized interaction operator $H_{int}^{(s)}(\tau)$ has the form

$$\left\langle \mathbf{p}_2 \mid H_{int}^{(s)}(\tau) \mid \mathbf{p}_1 \right\rangle = -2i\lambda\delta(\tau)\varphi^*(\mathbf{p}_2)\varphi(\mathbf{p}_1),$$

and hence the dynamics generated by this operator is equivalent to the dynamics governed by the Schrödinger equation with the separable potential

$$\left\langle \mathbf{p}_2 \mid H_I \mid \mathbf{p}_1 \right\rangle = \lambda\varphi^*(\mathbf{p}_2)\varphi(\mathbf{p}_1). \tag{29}$$

Solving Eq.(25) with the boundary condition (28), we easily get the well-known expression for the $T$ matrix in the separable-potential model

$$\left\langle \mathbf{p}_2 \mid T(z) \mid \mathbf{p}_1 \right\rangle = \varphi^*(\mathbf{p}_2)\varphi(\mathbf{p}_1) \left( \frac{1}{\lambda} - \int \frac{d^3k}{(2\pi)^3} \frac{|\varphi(\mathbf{k})|^2}{z - E_k} \right)^{-1}. \tag{30}$$

Ordinary quantum mechanics does not permit the extension of the above model to the case $\alpha \le \frac{1}{2}$. Indeed, in the case of such a high momentum behavior of the form factors $\varphi(\mathbf{p})$, the use of the interaction Hamiltonian given by (29) leads to the UV divergences, i.e., the integral in (30) is not convergent. We will now show that the generalized dynamical equation (10) allows one to extend this model to the case $-\frac{1}{2} < \alpha \le \frac{1}{2}$. Let us determine the class of the functions $f_1(z)$ and correspondingly the value of $\beta$ for which Eq.(25) has a unique solution having the asymptotic behavior (26). In the case $\alpha < \frac{1}{2}$, the function $t(z)$ given by Eq.(27) has the following behavior for $\mid z \mid \longrightarrow \infty$:

$$t(z) \xrightarrow[|z| \longrightarrow \infty]{} b_1(-z)^{\alpha - \frac{1}{2}} + b_2(-z)^{2\alpha - 1} + o(\mid z \mid^{2\alpha - 1}), \qquad (31)$$

where $b_1 = -4\pi \cos(\alpha \pi)(2\mu)^{\alpha - \frac{3}{2}}$ and $b_2 = -b_1^2 (M(0) + g_0^{-1})$ with

$$M(s) = \int \frac{d^3 k}{(2\pi)^3} \frac{\mid \varphi(\mathbf{k}) \mid^2 - \mid \mathbf{k} \mid^{-2\alpha}}{s - E_k}.$$

Here we restrict ourselves to the case where the function $\mid \varphi(\mathbf{k}) \mid^2 - \mid \mathbf{k} \mid^{-2\alpha}$ vanishes rapidly enough at infinity for the integral in this equation to be convergent. The parameter $b_1$ does not depend on $g_0$, i.e., on the value of the function $t(z)$ at the point $t = 0$. This means that all solutions of Eq.(25) have the same leading term in Eq.(31), and only the next-to-leading order term distinguishes the different solutions of this equation. Thus, in order to obtain a unique solution of Eq.(25), we must specify the first two terms in the asymptotic behavior of $t(z)$ for $\mid z \mid \longrightarrow \infty$. From this it follows that the functions $f_1(z)$ must be of the form

$$f_1(z) = b_1(-z)^{\alpha - \frac{1}{2}} + b_2(-z)^{2\alpha - 1},$$

and $\beta = 1 - 2\alpha$. Correspondingly, the functions $f(\tau)$ must be of the form

$$f(\tau) = a_1 \tau^{-\alpha - \frac{1}{2}} + a_2 \tau^{-2\alpha},$$

with $a_1 = -ib_1 \Gamma^{-1}(\frac{1}{2} - \alpha) \exp[i(-\frac{\alpha}{2} + \frac{1}{4})\pi]$, and the free parameters of the theory $a_2$ and $b_2$ are related by $a_2 = b_2 \Gamma^{-1}(1 - 2\alpha) \exp(-i\alpha \pi)$, where $\Gamma(z)$ is the gamma-function. This means that in the case $\alpha < \frac{1}{2}$ the generalized interaction operator must be of the form

$$\left\langle \mathbf{p}_2 \mid H_{int}^{(s)}(\tau) \mid \mathbf{p}_1 \right\rangle = \varphi^*(\mathbf{p}_2)\varphi(\mathbf{p}_1)\left( a_1 \tau^{-\alpha - \frac{1}{2}} + a_2 \tau^{-2\alpha} \right), \qquad (32)$$

i.e., must be nonlocal-in-time, and, as a consequence, the dynamics of the system is nonlocal in time. As it follows from Eqs.(24) and (27), in the case of such an interaction operator, the solution of Eq.(19) is of the form

$$\langle \mathbf{p}_2 \mid T(z) \mid \mathbf{p}_1 \rangle = N(z)\varphi^*(\mathbf{p}_2)\varphi(\mathbf{p}_1), \tag{33}$$

with

$$N(z) = g_0 \left( 1 - z g_0 \int \frac{d^3 k}{(2\pi)^3} \frac{\mid \varphi(\mathbf{k}) \mid^2}{(z - E_k) E_k} \right)^{-1} = \frac{b_1^2}{-b_2 + b_1(-z)^{\frac{1}{2}-\alpha} - M(z) b_1^2}.$$

Here we have taken into account that

$$g_0 = -b_1^2 \left( b_2 + b_1^2 M(0) \right)^{-1}.$$

By using Eqs.(15) and (33), we can construct the evolution operator

$$\begin{aligned}
\langle \mathbf{p}_2 \mid U(t, t_0) \mid \mathbf{p}_1 \rangle &= \langle \mathbf{p}_2 \mid \mathbf{p}_1 \rangle + \frac{i}{2\pi} \int_{-\infty}^{\infty} dx \\
&\times \frac{\exp[-i(z - E_{p_2})t] \exp[i(z - E_{p_1})t_0]}{(z - E_{p_2})(z - E_{p_1})} \\
&\times \varphi^*(\mathbf{p}_2)\varphi(\mathbf{p}_1) N(z),
\end{aligned} \tag{34}$$

where $z = x + iy$, and $y > 0$. The evolution operator $U(t, t_0)$ defined by Eq.(34) is a unitary operator satisfying the composition law (3), provided that the parameter $b_2$ is real.

The case $\alpha = \frac{1}{2}$ was considered in Ref.[12] where it has been shown that in this case the generalized interaction operator is of the form

$$\langle \mathbf{p}_2 \mid H_{int}^{(s)}(\tau) \mid \mathbf{p}_1 \rangle = -\frac{i}{2\pi} \varphi^*(\mathbf{p}_2)\varphi(\mathbf{p}_1) \int_{-\infty}^{\infty} dx \exp(-iz\tau) \left( \frac{b_1}{\ln(-z)} + \frac{b_2}{\ln^2(-z)} \right),$$

where $b_1 = -\frac{2\pi^2}{\mu}$. This means that in the case $\alpha = \frac{1}{2}$ the interaction is also nonlocal in time, and the dynamics is non-Hamiltonian. Thus, as we have seen, in the case $\alpha > \frac{1}{2}$ where the form factors meet the requirement of the ordinary quantum mechanics, the separable interaction may be only instantaneous and the dynamics is necessarily Hamiltonian. On the contrary, in the case $-\frac{1}{2} < \alpha < \frac{1}{2}$ (the restriction $\alpha > -\frac{1}{2}$ is necessary for the integral in (27) to be convergent) where the form factors have the "bad" high momentum behavior which within Hamiltonian dynamics gives rise to the ultraviolet divergences, the interaction is necessarily nonlocal in time, and hence the dynamics is non-Hamiltonian.

## Nucleon Dynamics in the EFT of Nuclear Forces

The technique of effective field theories is largely used in many branches of physics where a separation of scale exists. In low energy nuclear systems, the scale are, on one side, the low scales of the typical momentum of the process considered and the pion mass $m_\pi$, and, on the other side the higher scale associated with the chiral symmetry. The separation of scales produces a low energy expansion which allows one to make model independent predictions for low energy nuclear phenomena by using an effective Lagrangian including nuclens and pions as explicit degrees of freedom and all possible interactions that are consistent with the symmetries of QCD. This method, known as chiral perturbation theory, has been successfully applied to processes involving $0$ and $1$ nucleons [13,14]. In his pioneering work [1] Weinberg proposed to extend EFT methods to systems containing two or more nucleons. Because at low energy scales the momenta of the nucleons are small compared to their rest mess, the theory becomes nonrelativistic at leading order in the velocity expansion, with relativistic corrections included at higher orders. Thus the most general chirally invariant Lagrangian consists of contact interactions between nonrelativistic nucleons, and between nucleons and pions.

At very low energy, even the pion field may be integrated out. In this case we may consider an effective field theory consisting solely of nucleon fields. For this theory the chiral Lagrangian is reduced to the following Lagrangian

$$L_{NN} = N^+ i \partial_t N + N^+ \frac{\nabla^2}{2m} N - \frac{1}{2} C_S \left(N^+ N\right)^2 -$$
$$- \frac{1}{2} C_T \left(N^+ \sigma N\right)^2 - \frac{1}{4} C_2 (N^+ \nabla^2 N)(N^+ N) \qquad (35)$$
$$+ h.c. + \cdots,$$

where $\sigma$ are the Pauli matrices acting on spin indices, $m$ is the mass of nucleon, and $N$ denotes the nucleon field. The coefficients $C_S$ and $C_T$ are the couplings introduced by Weinberg [1].

Let us consider the two-nucleon system in the $^1S_0$ channel at very low energy. At extreme low energy we may restrict ourselves to leading order. At this order the effective Lagrangian takes the form

$$L_{NN} = N^+ i \partial_t N + -N^+ \frac{\nabla^2}{2m} N - \frac{1}{2} C_S \left(N^+ N\right)^2 - \frac{1}{2} C_T \left(N^+ \sigma N\right)^2. \qquad (36)$$

The two-nucleon $T$ matrix can be obtained by summing Feynman diagrams computed in this theory. Summing the bubble diagrams and absorbing the UV divergences in the renormalized parameter $C_R$ yield the following the $^1S_0$ channel two-nucleon $T$ matrix in the center-of-mass frame [15,16]:

$$\langle \mathbf{p}_2 \mid T^{(0)}(z) \mid \mathbf{p}_1 \rangle = \left[ \frac{1}{C_R} - \frac{m^{\frac{3}{2}}\sqrt{-z}}{4\pi} \right]^{-1}. \tag{37}$$

The $T$ matrix (37) does not satisfy the $LS$ equation. The above means that low energy nucleon dynamics is non-Hamiltonian, and the interaction which generates this dynamics cannot be parametrized by some interaction Hamiltonian. However, this result does not mean that the EFT is inconsistent with the basic principles of quantum mechanics because only the generalized dynamical equation must be satisfied in any case not the Schrödinger (LS) equation, and in Ref.[8] it has been shown that the $T$ matrix (37) satisfies the generalized dynamical equation with a nonlocal-in-time interaction operator. In other words, renormalization of the EFT results in the fact that the effective $NN$ interaction becomes nonlocal in time. To clarify this point let us consider another way of constructing the two nucleon $T$ matrix in the EFT approach.

The Weinberg program for low energy nucleon physics employs the analysis of time-ordered diagrams in ChPT to derive a $NN$ potential and then to use it in the full two-nucleon T matrix. As it follows from the Weinberg analysis [1], nucleon dynamics at very low energy in the $^1S_0$ channel should be governed by the potential

$$V(\mathbf{p}_2,\mathbf{p}_1) = C_0 + C_2(p_1^2 + p_2^2) + \cdots, \tag{38}$$

where $C_0 = C_S - 3C_T$. Obviously, this potential is singular, and hence the corresponding LS equation makes no sense without regularization and renormalization. Perform these procedures at leading order. In this case the $NN$ potential takes the form

$$V(\mathbf{p}_2,\mathbf{p}_1) = C_0. \tag{39}$$

Our goal is to show how regularization of the LS equation with this potential spreads the effective interaction in time. As we have noted, in order that the $NN$ interaction be instantaneous the T matrix should have the following high energy behavior

$$\langle \mathbf{p}_2 \mid T(z) \mid \mathbf{p}_1 \rangle \xrightarrow[|z|\to\infty]{} V(\mathbf{p}_2,\mathbf{p}_1). \tag{40}$$

Only in this case, as it follows from Eqs.(13) and (16), the interaction is instantaneous and the dynamics is governed by the Schrödinger (LS) equation. Let us regularize the singular potential (39) by using a momentum cut-off. In this case this potential is replaced by the regularized one

$$V_\Lambda(\mathbf{p}_2,\mathbf{p}_1) = f^*(p_2/\Lambda)C_0(\Lambda)f(p_1/\Lambda), \tag{41}$$

where the form factor $f(p/\Lambda)$ satisfies $f(0) = 1$ and falls off rapidly for $p/\Lambda > 0$. The form factor $f(p/\Lambda)$ composes the cut-off on the high momentum behavior of the amplitudes, and hence the LS equation with this such a potential is well defined and yields the following $T$ matrix

$$\langle \mathbf{p}_2 \mid T_\Lambda(z) \mid \mathbf{p}_1 \rangle = f^*(p_2/\Lambda) f(p_1/\Lambda)$$
$$\times \left( C_0^{-1}(\Lambda) - \int \frac{d^3k}{(2\pi)^3} \frac{\mid f(k/\Lambda) \mid^2}{z - E_k} \right)^{-1}. \tag{42}$$

Obviously the regularized $T$ matrix satisfies the asymptotic behavior

$$\langle \mathbf{p}_2 \mid T_\Lambda(z) \mid \mathbf{p}_1 \rangle \xrightarrow[\mid z \mid \to \infty]{} V_\Lambda(\mathbf{p}_2, \mathbf{p}_1), \tag{43}$$

which emphasized that the interaction is instantaneous. Let us define the renormalized value $C_R$ of $C_0$ as the value of the $T$ matrix $\langle \mathbf{p}_2 \mid T_\Lambda(z) \mid \mathbf{p}_1 \rangle$ at $z = \frac{p_1^2}{m} = \frac{p_2^2}{m} = 0$. That is

$$C_R^{-1} = C_0^{-1}(\Lambda) + \int \frac{d^3k}{(2\pi)^3} \frac{\mid f(k/\Lambda) \mid^2}{E_k}. \tag{44}$$

The value of the renormalized constant $C_R$ is fixed by the scattering length

$$a = m C_R / 4\pi.$$

Using Eq.(44), we may rewrite Eq.(43) as

$$\langle \mathbf{p}_2 \mid T_\Lambda(z) \mid \mathbf{p}_1 \rangle = \frac{f^*(p_2/\Lambda) f(p_1/\Lambda)}{C_R^{-1} - z \int \frac{d^3k}{(2\pi)^3} \frac{\mid f(k/\Lambda) \mid^2}{(z - E_k) E_k}}. \tag{45}$$

Thus, after renormalization the integral in the expression for the leading order two-nucleon $T$ matrix is effectively cut off, and hence at this stage the regularization may be removed. Letting $\Lambda \longrightarrow \infty$ in Eq.(45), we arrive at expression (37) for the two-nucleon $T$ matrix. As we see, for any finite $\Lambda$ the regularized $T$ matrix satisfies the LS equation and tends to the nonzero potential $V_\Lambda(\mathbf{p}_1, \mathbf{p}_2)$. On the other hand, $V_\Lambda(\mathbf{p}_1, \mathbf{p}_2)$ itself tends to zero as $\Lambda \longrightarrow \infty$. Thus there are no potentials that could produce the $T$ matrix of the form (37), because in the Hamiltonian formalism equality of a potential to zero means that there is no interaction in the system. From more general point of view provided by the GQD we see that the above does not lead to some problems and only means that low energy nucleon dynamics is governed by a nonlocal-in-time interaction and hence is non-Hamiltonian. In fact, from Eq. (37) it follows that the two-nucleon $T$ matrix has the following asymptotic behavior

$$\langle \mathbf{p}_2 \mid T^{(0)}(z) \mid \mathbf{p}_1 \rangle \xrightarrow[|z| \to \infty]{} -4\pi m^{-3/2}(-z)^{-1/2}$$
$$-16\pi^2 m^{-3} C_R^{-1}(-z)^{-1} + o(|z|^{-1}). \tag{46}$$

As it follows from Eqs.(11) and (16), such a behavior of the $T$ matrix corresponds to the following nonlocal-in-time interaction operator

$$\langle \mathbf{p}_2 \mid H_{int}^{(s)}(\tau) \mid \mathbf{p}_1 \rangle = \sqrt{\frac{16\pi}{m^3}} \exp(-i\pi/4)\left(\tau^{-\frac{1}{2}} + i\gamma\right), \tag{47}$$

where $\gamma = 4(\pi/m)^{3/2}\exp(-i\pi/4)(C_R)^{-1}$. The $T$ matrix (37) is the solution of the generalized dynamical equation with this interaction operator. Moreover, this operator is a particular case of the interaction operator in our model considered in the previous section (see, Eq.(32)).

The above shows that the problem of UV divergences is the cost of trying to extract from the Weinberg analysis of diagrams for the two-nucleon $T$ matrix in ChPT some effective Hamiltonian, while from this analysis it really follows that the effective $NN$ interaction is nonlocal in time. Let us now show that the GQD allows one to construct the $T$ matrix directly from the Weinberg analysis without resorting to some effective Hamiltonians or Lagrangians which give rise to UV divergences. The starting point for the Weinberg program is the assumption that in the nonrelativistic limit ChPT leads to low energy nucleon dynamics that is Hamiltonian and is governed by the Schrödinger equation. However, the fact that the chiral potentials constructed in this way are singular and lead to UV divergences means that this assumption has not proved correct. At the same time the GQD allows one to analyze the predictions of ChPT without making any preliminary assumptions about the character of low energy nucleon dynamics: This character should results from the analysis. Let us consider, for example, the two-nucleon system in the $^1S_0$ channel. At very low energy even the pion field can be integrated out, and the diagrams of the ChPT take the form of the diagrams being produced by the effective Lagrangian containing only contact interactions among nucleons and derivatives thereof. From the analysis of diagrams in this theory it follows that the two-nucleon $T$ matrix in the $^1S_0$ channel must be of the form

$$\langle \mathbf{p}_2 \mid T(z) \mid \mathbf{p}_1 \rangle = \sum_{n,m=0}^{\infty} p_2^{2n} p_1^{2m} t_{nm}(z), \tag{48}$$

where $\mathbf{p}_i$ is a relative momentum of nucleons. This expression is a chiral expansion of the $T$ matrix in powers of $Q/\Lambda$, where $Q$ is some low energy scale, and $\Lambda$ being the scale of the spontaneous symmetry violation. The terms $t_{nm}(z)$ are of order $|t_{nm}(z)| \sim O\left\{(Q/\Lambda)^{2(n+m)}\right\}$. From Eq.(48) it follows that the leading order $T$ matrix should be of the form

$$\left\langle \mathbf{p}_2 \mid T^{(0)}(z) \mid \mathbf{p}_1 \right\rangle = t_{00}(z),  \tag{49}$$

Thus ChPT results in the fact that the leading order contact component of the two-nucleon $T$ matrix is momentum-independent, i.e., is separable with the unite form factor. As has been shown, in the case of the bad UV behavior of the form factors, the dynamics of the system is non-Hamiltonian and is governed by the generalized dynamical equation with a nonlocal-in-time interaction operator whose form is completely determined by the behavior of the form factor. Obviously, the unite form factor has such a bad behavior (the $T$ matrix of the form (49) cannot satisfy the LS equation), and hence the effective operator of the $NN$ interaction must necessarily be nonlocal in time. As has been demonstrated above, this means that the $T$ matrix must be of the form (33), with $\alpha = 0$, and $\varphi(\mathbf{p}) = 1$

$$\left\langle \mathbf{p}_2 \mid T^{(0)}(z) \mid \mathbf{p}_1 \right\rangle = \frac{b_1^2}{-b_2 + b_1 \sqrt{-z}} = \left[ \frac{1}{C_R} - \frac{m\sqrt{-zm}}{4\pi} \right]^{-1},  \tag{50}$$

with $b_1 = -4\pi m^{-3/2}$ and $C_R = -b_1^2 / b_2$. Correspondingly the interaction operator parametrizing the $NN$ interaction at leading order must be of the form (49). Thus the requirement that the leading order two-nucleon $T$ matrix be of the form (49) and satisfy the generalized dynamical equation determines it up to one arbitrary parameter $C_R$. On the other hand, the above requirements are equivalent to those that the theory satisfy the basic principles of quantum mechanics and be consistent with the symmetries of QCD. This means that the effective theory of nuclear forces can be constructed as an inevitable consequence of the basic principles of quantum mechanics and the symmetries of the underlying quark-gluon physics. In contrast with the EFT approach to the theory of nuclear forces, where expression (37) is obtained by summing bubble diagrams and performing regularization and renormalization procedures, in this case the theory leads to the same results being finite at all the stages. At leading order of the pionless theory the $NN$ interaction is parametrized by the nonlocal-in-time interaction operator (47). This operator is well defined, and the generalized dynamical equation with this operator does not requires regularization and renormalization.

An advantage of the fact that within GQD the $T$ matrix is well defined is that one may use it in Eq. (9) for constructing the evolution operator

$$< \mathbf{p}_2 \mid U(t, t_0) \mid \mathbf{p}_1 > = < \mathbf{p}_2 \mid \mathbf{p}_1 > + \frac{i}{2\pi} \int_{-\infty}^{\infty} dx$$
$$\times \frac{\exp[-i(z - E_{p_2})t] \exp[i(z - E_{p_1})t_0]}{(z - E_{p_2})(z - E_{p_1})} \left[ \frac{1}{C_R} - \frac{m^{\frac{3}{2}}\sqrt{-z}}{4\pi} \right]^{-1},  \tag{51}$$

where $z = x + iy$, and $y > 0$. In contrast, the standard methods of the EFT approach to the theory of nuclear forces allow one to calculate only scattering amplitudes but not to construct the evolution operator. Note in this connection, that the $S$ matrix is not everything. For example, at finite temperature there is no $S$ matrix because particles cannot get out to infinite distances from a collision without bumping into things.

## THE HILBERT SPACE OF NUCLEON STATES

The main lesson we have learned from the analysis of the previous section is that low energy dynamics of nucleons in effective theory of nuclear forces is not governed by the Schrödinger equation: It is governed by the generalized dynamical equation with a nonlocal-in-time interaction operator when this equation cannot be reduced to the Schrödinger equation. This implies, that as it follows from Stone's theorem, the group of the evolution operators is not weekly continuous, i.e., condition (6) is not valid for all vectors belonging to the Hilbert space. At the same time, as it follows from the requirement of the physical continuity, this condition must be satisfied for physically realizable states such as states for which $\left\| H_0 \mid \psi \right\rangle \right\| < \infty$. From this in turn it follows that condition (6) has to be valid for all vectors $\mid \psi \rangle \in \overline{D(H_0)}$, where $\overline{D(H_0)}$ denotes the closure of $D(H_0)$. This means that condition (6) must be valid for all vectors belonging to the space $L^2(M)$, because $D(H_0)$ is dense in this space (we consider spinless particles). On the other hand, in the case where the dynamics of a quantum system is non-Hamiltonian, there must be states for which the continuity condition (6) is not satisfied. From this it follows that the Hilbert space describing states of such a system cannot be realized as the space $L^2(M)$. In other words, in this case states of the quantum system cannot be described by wave functions being the squire integrable functions of momenta of particles. Thus the Hilbert space describing states of nucleons cannot be realized as the space $L^2(M)$. Let us now investigate this situation by using the model we have considered above. Since the same problem should arise in any theory where the fundamental interaction is nonlocal in time, we will not restrict ourselves to the particular case where $\varphi(\mathbf{p}) = 1$ corresponding to low energy nucleon dynamics.

In the case of this model the operator $\tilde{S}(t_2, t_1)$ can be represented in the form

$$\tilde{S}(t_2, t_1) = D(t_2, t_1)\tilde{F}(t_2 - t_1),\tag{52}$$

where $\tilde{F}(\tau)$ is a function of $\tau$, and $D(t_2, t_1) = \exp(iH_0 t_2) \mid \varphi \rangle \langle \varphi \mid \exp(-iH_0 t_1)$ is the operator-valued distribution such that

$$\langle \mathbf{p}_2 \mid D(t_2, t_1) \mid \mathbf{p}_1 \rangle = \exp(iE_{p_2} t_2) \langle \mathbf{p}_2 \mid \varphi \rangle \langle \varphi \mid \mathbf{p}_1 \rangle \exp(-iE_{p_1} t_1).\tag{53}$$

Substituting Eq.(52) into the generalized dynamical equation (10), for $\tilde{F}(\tau)$, we get

$$(t_2 - t_1)\tilde{F}(t_2 - t_1) = \int_{t_1}^{t_2} dt_4 \int_{t_1}^{t_4} dt_3 (t_4 - t_3) \int \frac{d^3 k}{(2\pi)^3} \exp[-iE_k(t_4 - t_3)]$$
$$\times \langle \varphi \mid \mathbf{k} \rangle \langle \mathbf{k} \mid \varphi \rangle \tilde{F}(t_2 - t_4) \tilde{F}(t_3 - t_1). \tag{54}$$

Let us examine this equation in the limit $t_2 \longrightarrow t_1$. For this we have to change the variables: $\tau_i = \theta_i / \nu^2$, $\mathbf{k} = \nu \mathbf{q}_\nu$, $i = 1, 2, 3, 4$. In this way, Eq.(54) can be rewritten in the form

$$(\theta_2 - \theta_1)\tilde{F}(\theta_2 / \nu^2 - \theta_1 / \nu^2) =$$
$$= \nu^{-4} \int_{\theta_1}^{\theta_2} d\theta_4 \int_{\theta_1}^{\theta_4} d\theta_3 (\theta_4 - \theta_3) \int \frac{d^3 q_\nu}{(2\pi)^3} \exp[-iE_{q_\nu}(\theta_4 - \theta_3)] \tag{55}$$
$$\times \langle \varphi \mid \mathbf{q}_\nu \rangle \langle \mathbf{q}_\nu \mid \varphi \rangle \tilde{F}(\theta_2 / \nu^2 - \theta_4 / \nu^2) \tilde{F}(\theta_3 / \nu^2 - \theta_1 / \nu^2),$$

where $\mid \mathbf{q}_\nu \rangle = \nu^{\frac{3}{2}} \mid \mathbf{k} \rangle$. The vectors $\mid \mathbf{q}_\nu \rangle$ are the basis vectors for which the completeness conditions reads

$$\int \frac{d^3 q_\nu}{(2\pi)^3} \mid \mathbf{q}_\nu \rangle \langle \mathbf{q}_\nu \mid = 1.$$

From Eq.(55) it follows that the leading order term in the asymptotic behavior of $\tilde{F}(\tau)$ for $\tau \to 0$ is of the form $a_1 \tau^{-\alpha - \frac{1}{2}}$ with

$$a_1 = (\theta_2 - \theta_1)^{\frac{3}{2} - \alpha}$$
$$\times \lim_{\nu \to \infty} \Big[ \int_{\theta_1}^{\theta_2} d\theta_4 (\theta_2 - \theta_4)^{-\alpha - \frac{1}{2}} \int_{\theta_1}^{\theta_4} d\theta_3 (\theta_3 - \theta_1)^{-\alpha - \frac{1}{2}} (\theta_4 - \theta_3) \times$$
$$\times \int \frac{d^3 q_\nu}{(2\pi)^3} \exp[-iE_{q_\nu}(\theta_4 - \theta_3)] \langle \varphi_\nu \mid \mathbf{q}_\nu \rangle \langle \mathbf{q}_\nu \mid \varphi_\nu \rangle \Big]^{-1} \tag{56}$$
$$= 4\pi \cos(\alpha\pi) m^{\alpha - \frac{3}{2}} \Gamma^{-1}(\frac{1}{2} - \alpha) \exp\Big[\frac{i\pi}{2}(1/2 - \alpha)\Big],$$

where $\mid \varphi_\nu \rangle = \nu^{-3/2 + \alpha} \mid \varphi \rangle$. Here we have taken into account that from the asymptotic behavior (23) of the form factor $\varphi(\mathbf{p}) = \langle \varphi \mid \mathbf{p} \rangle$ it follows that

$$\langle \varphi_\nu \mid \mathbf{q}_\nu \rangle = \mid \mathbf{q}_\nu \mid^{-\alpha} + o\Big(\mid \mathbf{q}_\nu \mid^{-1 + \alpha}\Big), \quad \nu \to \infty, \tag{57}$$

Thus the leading order term in the asymptotic behavior of $\tilde{F}(\tau)$ for $\tau \to 0$ is uniquely determined by the value of the parameter $\alpha$ characterizing the high momentum behavior of the form factor $\psi(\mathbf{p})$. It is not difficult to verify now that the next-to-leading order term in the asymptotic behavior of $\tilde{F}(\tau)$ is $a_2\tau^{-2\alpha}$. In contrast with $a_1$, the parameter $a_2$ is arbitrary. In fact, substituting $\tilde{F}(\tau) = a_1\tau^{-\alpha-\frac{1}{2}} + a_2\tau^{-2\alpha} + o(\tau^{-2\alpha})$ into the right-hand side of Eq.(54) yields $a_1\tau^{-\alpha-\frac{1}{2}} + a_2\tau^{-2\alpha} + o(\tau^{-2\alpha})$. Thus, up to next-to-leading order Eq.(54) is fulfilled for any $a_2$. The above means that only the next-to-leading order term in the asymptotic behavior of $\tilde{F}(\tau)$ for $\tau \to 0$ distinguishes the different solutions of the generalized dynamical equation, and $a_2$ is a free parameter of the theory. Thus, in order to obtain a unique solution of Eq.(54) we must specify the leading and next-to-leading terms in the asymptotic behavior of $\tilde{F}(\tau)$, and hence the generalized interaction operator must be of the form (47). On the other hand, such a form of this operator implies that knowing these terms is sufficient to obtain the higher order terms in the asymptotic behavior of $\tilde{F}(\tau)$. Let us demonstrate this fact by using the example of the next-to-next-to-leading order term. It can be shown that this term is of the form $a_3\tau^{-3\alpha+\frac{1}{2}}$. Substituting $\tilde{F}(\tau) = a_1\tau^{-\alpha-\frac{1}{2}} + a_2\tau^{-2\alpha} + a_3\tau^{-3\alpha+\frac{1}{2}} + o(\tau^{-3\alpha+\frac{1}{2}})$ into Eq.(54), for the parameter $a_3$, we get

$$
\begin{aligned}
a_3 = 3(a_2)^2 \lim_{\nu\to\infty} (\theta_2-\theta_1)^{3\alpha-\frac{3}{2}} \int_{\theta_1}^{\theta_2} d\theta_4 \int_{\theta_1}^{\theta_4} d\theta_3 (\theta_2-\theta_4)^{-2\alpha} \\
\times (\theta_3-\theta_1)^{-2\alpha}(\theta_4-\theta_3) \int \frac{d^3q_\nu}{(2\pi)^3} \exp[-iE_{q_\nu}(\theta_4-\theta_3)] \langle \varphi_\nu \mid \mathbf{q}_\nu \rangle \langle \mathbf{q}_\nu \mid \varphi_\nu \rangle
\end{aligned}
\tag{58}
$$

Taking the limit in Eq.(58) and performing integrations yield

$$
a_3 = \frac{(a_2)^2}{a_1} \frac{\Gamma^2(1-2\alpha)}{\Gamma(\frac{1}{2}-\alpha)\Gamma(\frac{3}{2}-3\alpha)}.
$$

In this way one can obtain the higher order terms in the asymptotic behavior of $\tilde{F}(\tau)$ for $\tau \longrightarrow 0$ and than construct the operator $\tilde{S}(t_2,t_1)$ for any $t_1$ and $t_2$.

It is easy to see that, in the case of Eqs.(56) and (58) determining the asymptotic behavior of the operator $\tilde{S}(t_2,t_1)$ for $t_2 \longrightarrow t_1$ we deal with intermediate states with infinite momentum. Note in this connection that for any vector $\mid \psi \rangle$ belonging to the Hilbert space $H$

$$
\mid \psi \rangle = \int \frac{d^3k}{(2\pi)^3} \psi(\mathbf{k}) \mid \mathbf{k} \rangle,
$$

with $\psi(\mathbf{k}) = \langle \mathbf{k} \mid \psi \rangle$, one can construct another vector

$$|\psi_\nu\rangle = \int \frac{d^3k}{(2\pi)^3} \psi(\mathbf{k}/\nu)\nu^{\frac{3}{2}} |\mathbf{k}\rangle,$$

that represents the same physical state, if we scale $\mathbf{k} \rightarrow \nu\mathbf{k}$, i.e.,

$$|\psi_\nu\rangle = \int \frac{d^3q_\nu}{(2\pi)^3} \psi(\mathbf{q}_\nu) |\mathbf{q}_\nu\rangle,$$

where $\mathbf{q}_\nu = \mathbf{k}/\nu$. Varying the parameter $\nu$, we get a set of vectors having the same norm

$$\left\| \psi_\nu \right\| = \left( \int \frac{d^3q_\nu}{(2\pi)^3} \psi^*(\mathbf{q}_\nu)\psi(\mathbf{q}_\nu) \right)^{\frac{1}{2}}.$$

Each of the vectors $|\psi_\nu\rangle$ belongs to the Hilbert space $H$ even when the parameter $\nu$ is letting to infinity. Let us consider, for example, the set of vectors

$$|\psi_\nu^{(0)}\rangle = \int \frac{d^3q_\nu}{(2\pi)^3} \psi_{q_0}(\mathbf{q}_\nu) |\mathbf{q}_\nu\rangle, \tag{59}$$

where $\psi_{q_0}(\mathbf{q}_\nu)$ is zero everywhere outside the subset $\Delta(q_0)$ $(E_{q_0}\nu^2 \leq E_k \leq E_{q_0}\nu^2 + \varepsilon E_{q_0}\nu^2)$ of the spectrum of $H_0$. Vector (59) is a eigenvector of the projection operator $P_{\Delta(q_0)}$ on the subset $\Delta(q_0)$. The projection operators are defined for any subsets even when their location tend to the infinite part of the spectrum. Correspondingly eigenvectors of this operator for any subset of spectrum $H_0$ belong to the Hilbert space. In the case where the location of a subset tends to infinite part of the spectrum, corresponding eigenvectors represent states with infinite energy. For describing such states we have to make a change the scale by letting $\nu$ to infinity.

For any $\nu$ the vectors $|\psi_\nu\rangle$ can be represented as vectors belonging to the Hilbert space $L^2(M_\nu)$ of square integrable functions $\psi(\mathbf{q}_\nu)$, where $M_\nu$ denotes the momentum-space corresponding to the scale $\nu$. Obviously, for any finite $\nu$ the spaces $L^2(M_\nu)$ coincides each with other, since $\psi(\mathbf{q}_\nu) = \psi'(\mathbf{k}) = \psi(\mathbf{k}/\nu)$. This, of course, is not true for the space $L^2(M_\infty)$ of square integrable functions $\psi(\mathbf{q}_{in})$, with $\mathbf{q}_{in}$ being the momentum in the case when the scale tends to infinity. Since for any $|\mathbf{k}| < \infty$

$$\lim_{\nu \to \infty} \langle \mathbf{k} \mid \psi_\nu \rangle = \lim_{\nu \to \infty} \int \frac{d^3 q_\nu}{(2\pi)^3} \psi(\mathbf{q}_\nu) \langle \mathbf{k} \mid \mathbf{q}_\nu \rangle$$

$$= \lim_{\nu \to \infty} \int \frac{d^3 q_\nu}{(2\pi)^3} \psi(\mathbf{q}_\nu) \langle \mathbf{q}'_\nu \mid \mathbf{q}_\nu \rangle \nu^{-\frac{3}{2}} \qquad (60)$$

$$= \lim_{\nu \to \infty} \psi(\mathbf{k}/\nu) \nu^{-\frac{3}{2}} = 0,$$

any vector $\mid \psi_{in} \rangle \in L^2(M_\infty)$ is orthogonal to all vectors of $L^2(M_1)$. Thus the Hilbert space describing states of the system under study is

$$H = H_p \bigoplus H_{in}, \qquad (61)$$

where $H_p$ is the space that can be realized as the space $L^2(M_1)$, and $H_{in}$ is the space that can be realized as the space $L^2(M_\infty)$. Vectors belonging to the subspace $H_{in}$ represent states with infinite energy and are not physically realizable. The above means that in the case where the interaction is nonlocal in time we have to deal with the two type of the wave functions $\psi(\mathbf{k})$ and $\psi(\mathbf{q}_{in})$ which correspond to disparate energy scales. The physical reasons for this are quite obvious. A nonlocality in time of the interaction in a quantum system is a result of integrating out some degrees of freedom. In order that nevertheless such a reduced quantum system could be considered as closed, i.e., the evolution of the system be unitary, the above degrees of freedom should correspond to infinite energy scale. Thus the fact that the quantum system should be described by the Hilbert space (61) is a consequence of the fact that there are two well separated energy scales relevant for the problem under study.

## Non-Hamiltonian Character of Nucleon Dynamics

By using the results obtained in the previous section, let us now investigate the character of low energy nucleon dynamics. First of all, we will show that the evolution operator describing the evolution of a nucleon system is not weekly continuous. As it follows from Eqs.(8) and (9), the operator $R(t,t_0)$ defined by Eq. (4) can be represented in the form

$$R(t,t_0) = -i \int_{t_0}^{t} dt_2 \int_{t_0}^{t_2} dt_1 \times \exp(iH_0 t_2) \tilde{T}(t_2 - t_1) \exp(-iH_0 t_1). \qquad (62)$$

For the matrix element $\langle \psi_\nu^{(2)} \mid R(t,0) \mid \psi_\nu^{(1)} \rangle$, we have

$$\left\langle \psi_{\nu}^{(2)} \mid R(t,0) \mid \psi_{\nu}^{(1)} \right\rangle =$$

$$-i\int_{0}^{t} dt_2 \int_{0}^{t_2} dt_1 \int \frac{d^3 k'}{(2\pi)^3} \int \frac{d^3 k}{(2\pi)^3} \exp(iE_k t_2) \exp(-iE_k t_1) \left\langle \psi_{\nu}^{(2)} \mid \mathbf{k}' \right\rangle \quad (63)$$

$$\times \left\langle \mathbf{k}' \mid \tilde{T}(t_2 - t_1) \mid \mathbf{k} \right\rangle \left\langle \mathbf{k} \mid \psi_{\nu}^{(1)} \right\rangle.$$

Taking into account Eqs. (9), (11) and (32), and letting $\nu$ to infinity, we get

$$\left\langle \psi_{in}^{(2)} \mid R(t,0) \mid \psi_{in}^{(1)} \right\rangle = \lim_{\nu \to \infty} \left\langle \psi_{\nu}^{(2)} \mid R(t,0) \mid \psi_{\nu}^{(1)} \right\rangle =$$

$$-ia_1 \lim_{\nu \to \infty} \int_{0}^{t\nu^2} d\theta_2 \int_{0}^{\theta_2} d\theta_1 \int \frac{d^3 q'_{\nu}}{(2\pi)^3} \int \frac{d^3 q_{\nu}}{(2\pi)^3} \exp(iE_{q'_{\nu}} \theta_2)$$

$$\times \exp(-iE_{q_{\nu}} \theta_1) \psi_2^*(\mathbf{q}'_{\nu}) \psi_1(\mathbf{q}_{\nu}) \nu^{-1} \mid \nu \mathbf{q}_{\nu} \mid^{-\alpha} \mid \nu \mathbf{q}'_{\nu} \mid^{-\alpha} \left( \frac{\theta_2 - \theta_1}{\nu^2} \right)^{-\frac{1}{2} - \alpha},$$

where $\theta_i = t_i \nu^2$, $\psi_i(\mathbf{q}_{\nu}) = \left\langle \mathbf{q}_{\nu} \mid \psi_{\nu}^{i} \right\rangle$, $i = 1, 2$. From this it follows that in the limit $\nu \longrightarrow \infty$ the matrix elements $\left\langle \psi_{\nu}^{(2)} \mid R(t,0) \mid \psi_{\nu}^{(1)} \right\rangle$ are scale invariant, and we have

$$\left\langle \psi_{in}^{(2)} \mid R(t,0) \mid \psi_{in}^{(1)} \right\rangle = -ia_1 \int_{0}^{\infty} d\theta_2 \int_{0}^{\theta_2} d\theta_1 \int \frac{d^3 q'}{(2\pi)^3} \int \frac{d^3 q}{(2\pi)^3} \exp(iE_q \theta_2)$$

$$\times \exp(-iE_q \theta_1) \psi_2^*(\mathbf{q}') \psi_1(\mathbf{q}) \mid \mathbf{q} \mid^{-\alpha} \mid \mathbf{q}' \mid^{-\alpha} (\theta_2 - \theta_1)^{-\frac{1}{2} - \alpha},$$

where we denote momenta corresponding to the infinite scale by $\mathbf{q}$.

Thus values of the matrix elements $\left\langle \psi_{in}^{(2)} \mid R(t,0) \mid \psi_{in}^{(1)} \right\rangle$ are independent of $t$. Obviously these values are not zero for all wave functions $\psi_{in}(\mathbf{q}_{\nu})$. This means that in the general case the amplitudes $\left\langle \psi_{in}^{(2)} \mid R(t,0) \mid \psi_{in}^{(1)} \right\rangle$ are not continuous at $t = 0$ since, as it follows from the definition, $R(0,0) = 0$. From this in turn it follows that the evolution operator (34) is not weakly continuous. On the other hand one can show that for the physically realizable states $\mid \psi_1 \rangle$ and $\mid \psi_2 \rangle$ the matrix elements $\left\langle \psi_2 \mid U(t,0) \mid \psi_1 \right\rangle$ tend to $\left\langle \psi_2 \mid \psi_1 \right\rangle$ as $t \to 0$ and hence condition (6) is not violated. Nevertheless, the evolution operator $V(t) = U_s(t,0)$ is not continuous and hence the group of these operators has no infinitesimal generator in this case. This means that in this case the time evolution of a state vector is not governed by the Schrödinger equation. The physical cause of the discontinuity of the matrix elements of the operator $R(\delta t, 0)$ for states with infinite energy was explained above. The fact that in principle for describing the evolution of a quantum system during infinitesimal time intervals one has to take into account the intermediate states with infinite energy is a consequence of the principle of uncertainty. In the case of Hamiltonian dynamics, nevertheless, such states give no contributions because the matrix elements $\langle \mathbf{p}_2 \,|\, U(t_2, t_1) \,|\, \mathbf{p}_1 \rangle$ vanish sufficiently fast when momenta tend to infinity. In this case the states belonging to the space $H_{in}$ do not manifest themselves in any way, and the dynamics of the system can be described in terms of the space $H_p$. However, in general the matrix elements $\langle \mathbf{p}_2 \,|\, U(t_2, t_1) \,|\, \mathbf{p}_1 \rangle$ may have such a high momentum behavior that one cannot ignore the space $H_{in}$ in describing the dynamics of a quantum system, and in this case the dynamics is non-Hamiltonian.

From the above analysis it follows that in the case where the interaction in a quantum system is nonlocal in time, one has to take into account states belonging to $H_{in}$ in describing the time evolution of the system. On the other hand, the possibility to find the system at time $t$ in a state $|\psi_{in}\rangle \in H_{in}$, if initially at time $t_0$ the state of the system was physically realizable, must be zero. This means that the matrix element $\langle \psi_{in} \,|\, U(t,0) \,|\, \psi_1 \rangle$ with $|\psi_1\rangle \in H_p$ has to be zero for any time $t$. Let us show that this really takes place. For the above matrix element we can write

$$\langle \psi_{in} \,|\, U(t,0) \,|\, \psi_1 \rangle =$$

$$\lim_{\nu \to \infty} \int_0^t dt_2 \int_0^{t_2} dt_1 \int \frac{d^3k'}{(2\pi)^3} \int \frac{d^3k}{(2\pi)^3} \exp(iE_k t_2) \exp(-iE_{k'} t_1) \psi^*(\mathbf{k}/\nu) \nu^{-\frac{3}{2}} \varphi^*(\mathbf{k})$$

$$\times \langle \mathbf{k} \,|\, \tilde{T}(t_2 - t_1) \,|\, \mathbf{k}' \rangle \varphi(\mathbf{k}') \psi_1(\mathbf{k}') =$$

$$a_1 \lim_{\nu \to \infty} \int_0^{t\nu^2} d\theta_2 \int_0^{\theta_2} d\theta_1 \int \frac{d^3 q_\nu}{(2\pi)^3} \int \frac{d^3 k'}{(2\pi)^3} \exp(iE_{q_\nu} \theta_2)$$

$$\times (\theta_2 - \theta_1)^{-\frac{1}{2} - \alpha} \,|\, \mathbf{q}_\nu \,|^{-\alpha} \psi^*(\mathbf{q}_\nu) \psi_1(\mathbf{k}') \nu^{-\frac{3}{2}} = 0.$$

Thus, if at time $t_0$ the state of the system was physically realizable, the probability to find the system at time $t$ in any state $|\psi\rangle \in H_{in}$ is zero, despite the fact that one cannot ignore the subspace $H_{in}$. This means that the states $|\psi\rangle \in H_{in}$ are not observable. This is not at variance with the fact that one have to take into account the space $H_{in}$ in describing the time evolution of the system, because intermediate states belonging to $H_{in}$ are responsible for validity of the composition law (3) for infinitesimal time intervals. In order to demonstrate this fact, let us rewrite Eq.(3) in terms of the operator $R(t_2, t_1)$:

$$\langle \psi_2 \mid R(t,0) \mid \psi_1 \rangle = \langle \psi_2 \mid R(t,t') \mid \psi_1 \rangle + \langle \psi_2 \mid R(t',0) \mid \psi_1 \rangle$$

$$+ \int \frac{d^3k}{(2\pi)^3} \langle \psi_2 \mid R(t,t') \mid \mathbf{k} \rangle \langle \mathbf{k} \mid R(t',0) \mid \psi_1 \rangle, \tag{64}$$

$$t > t' > 0.$$

Taking into account Eqs. (9), (11), and (32) in the limit $t \longrightarrow t_0$ this equation can be written in the form

$$\nu^{-3+2\alpha} \langle \psi_2 \mid \tilde{R}(\theta,0) \mid \psi_1 \rangle =$$

$$\nu^{-3+2\alpha} \langle \psi_2 \mid \tilde{R}(\theta,\theta') + \tilde{R}(\theta',0) + i \int \frac{d^3 q_\nu}{(2\pi)^3} \tilde{R}(\theta,\theta') \mid \mathbf{q}_\nu \rangle \langle \mathbf{q}_\nu \mid \tilde{R}(\theta',0) \mid \psi_1 \rangle \tag{65}$$

$$+ o(\nu^{-3+2\alpha}), \quad \nu \longrightarrow \infty,$$

with

$$\tilde{R}(\theta,\theta') = -ia_1 \int_{\theta'}^{\theta} d\theta_2 \int_{\theta'}^{\theta_2} d\theta_1 (\theta_2 - \theta_1)^{-1/2-\alpha} \exp\left(\frac{iH_0\theta}{\nu^2}\right) \mid \varphi \rangle \langle \varphi \mid \exp\left(\frac{-iH_0\theta'}{\nu^2}\right).$$

From Eq.(57) it follows that $\nu^{-3/2+\alpha} \langle \psi \mid \tilde{R}(\theta_2,\theta_1) \mid \mathbf{q}_\nu \rangle$ and $\nu^{-3/2+\alpha} \langle \mathbf{q}_\nu \mid \tilde{R}(\theta_2,\theta_1) \mid \psi \rangle$ tend to finite nonzero limits as $\nu \longrightarrow \infty$. Thus for infinitesimal time intervals the composition law (3) reads

$$\langle \psi_2 \mid \tilde{R}(\theta,\theta_0) \mid \psi_1 \rangle =$$

$$\langle \psi_2 \mid \tilde{R}(\theta,\theta') \mid \psi_1 \rangle + \langle \psi_2 \mid \tilde{R}(\theta',\theta_0) \mid \psi_1 \rangle + \tag{66}$$

$$\lim_{\nu \to \infty} \nu^{-3+2\alpha} \int \frac{d^3 q_\nu}{(2\pi)^3} \langle \psi_2 \mid \tilde{R}(\theta,\theta') \mid \mathbf{q}_\nu \rangle \langle \mathbf{q}_\nu \mid \tilde{R}(\theta',\theta_0) \mid \psi_1 \rangle.$$

This equation proofs that indeed intermediate states belonging to $H_{in}$ are responsible for validity of the composition law for infinitesimal time intervals. However, this true only in the case $\alpha < \frac{1}{2}$ where the interaction is nonlocal in time, and hence the form factor $\varphi(\mathbf{p})$ has the "bad" UV behavior. In the case where the interaction operator is of the form (13) and, as a result, the dynamics of a quantum system is Hamiltonian, the states $\mid \psi \rangle \in H_{in}$ do not play any role in the time evolution of the system even in the infinitesimal neighborhood of the point $t = 0$. In fact, in this case from Eq.(62), for $t \longrightarrow 0$, we get

$$\langle \psi_2 \mid R(t_2,t_1) \mid \psi_1 \rangle = \nu^{-2} \langle \psi_2 \mid \tilde{R}(\theta_2,\theta_1) \mid \psi_1 \rangle + o(\nu^{-2}), \quad \nu \longrightarrow \infty, \tag{67}$$

where

$$\left\langle n_2 \mid \tilde{R}(\theta_2, \theta_1) \mid n_1 \right\rangle = \int_{\theta_1}^{\theta_2} d\theta' \left\langle n_2 \mid H_I(\theta'/\nu) \mid n_1 \right\rangle \tag{68}$$

Obviously Eq.(64) is valid in the limit $t \longrightarrow 0$ for any interaction Hamiltonian satisfying the ordinary requirements of quantum mechanics, and in this case only the first two terms on the right-hand side of (64) are relevant for $t \longrightarrow 0$. As a result, in the case of Hamiltonian dynamics, the Hilbert space of states of spinless particles can be realized as the space $L^2(M_1)$. The manifold of the physically realizable states, being the domain of $H_0$, is dense in this Hilbert space, and the evolution operator is strongly continuous. On the contrary, in the case where the interaction generating the dynamics of a system is nonlocal in time, and the interaction operator is not of the form (13), in the limit $t_2 \to t_1$ the matrix elements $\left\langle \psi_2 \mid \tilde{R}(t_2, t_1) \mid \psi_1 \right\rangle$ do not behave like (67). This means that in order that Eq.(64) be valid in this limit the contribution from the third term on the right-hand side of Eq.(64)

$$i \int \frac{d^3k}{(2\pi)^3} \left\langle \psi_2 \mid R(t, t') \mid \mathbf{k} \right\rangle \left\langle \mathbf{k} \mid R(t', 0) \mid \psi_1 \right\rangle$$

must be of the same order as that from the first two terms. However, for this in the high momentum limit $\left\langle \mathbf{k}_2 \mid R(t_2, t_1) \mid \mathbf{k}_1 \right\rangle$ must not vanish as fast as in the case of Hamiltonian dynamics, i.e., must have the "bad" UV behavior. In fact, in the limit $t \longrightarrow 0$ the main contribution to the above term comes from the intermediate states with infinite energies. This fact explains the above mentioned one-to-one correspondence between the character of the dynamics of a quantum system and the high momentum behavior of the matrix elements of the evolution operator.

The fact that, despite the subspace $H_p$ is invariant for the evolution operator, the description of the evolution of a system whose dynamics is generated by a nonlocal in time interaction cannot be reduced to this subspace can also be illustrated by using the following example. Let us consider the operator $V_p(t) = PV(t)P$, where $V(t)$ is evolution operator in the Schrödinger picture describing the dynamics generated by a nonlocal-in-time interaction operator (32), and $P$ is the projection operator on the subspace $H_p$. Assume that the group of the operators $V_p(t)$ has a self-adjoint infinitesimal generator $A$. Then, for $|\psi\rangle \in D(A)$, we have

$$\frac{V_p(t) \mid \psi \rangle - \mid \psi \rangle}{t} \xrightarrow[t \to 0]{} -iA \mid \psi \rangle.$$

From this it follows that $A = H_0 + A_1$, with $A_1 = -\lim_{t \to 0} \left( \frac{1}{t} \exp(-iH_0 t) R(t, 0) \right)$. By using Eq.(63), for $|\psi\rangle \in D(H_0)$ and $|\mathbf{k}| < \infty$, we get

$$\langle \mathbf{k} \mid A_1 \mid \psi \rangle = i \lim_{t \to 0} \left( \frac{1}{t} \int_0^t dt_2 \int_0^{t_2} dt_1 \right.$$

$$\times \int \frac{d^3 k_1}{(2\pi)^3} \exp[iE_k(t_2 - t)] \exp(-iE_{k_1} t_1) \varphi^*(\mathbf{k}) \varphi(\mathbf{k}_1) \langle \mathbf{k} \mid \tilde{T}(t_2 - t_1) \mid \mathbf{k}_1 \rangle \psi(\mathbf{k}_1) \Big)$$

$$= ia_1 C \varphi^*(\mathbf{k}) \lim_{t \to 0} \left( \frac{1}{t} \int_0^t dt_2 \int_0^{t_2} dt_1 (t_2 - t_1)^{-\frac{1}{2} - \alpha} \right) = 0,$$

where $C = \int \frac{d^3 k_1}{(2\pi)^3} \varphi(\mathbf{k}_1) \psi(\mathbf{k}_1)$. Hence $A = H_0$, i.e., the infinitesimal generator of the group of the operators $V_p(t)$ is equal to the free Hamiltonian. This means that the dynamics of the system cannot be restricted to the space $H_p$.

Let us now consider the physical aspects of the problem. As has been stated, integrating out certain degrees of freedom results in the fact that the effective action in a quantum system becomes nonlocal in time. This nonlocality does not give rise to a loss of probability from the system only if the energy scale $\Lambda_\infty$ at which these degrees of freedom are observable is sufficiently large and, from the point of view of the low energy theory, may be considered as infinite. In the EFT of nuclear forces the degrees of freedom integrated out in the description of low energy nucleon dynamics are antinucleons, heave vector mesons, $\Delta'$s, quarks and gluons (in the pionless theory pions are also integrated out). The energy scale associated with these degrees of freedom may be regarded as infinite, and, as a result, they are not observable in low energy regime. On the other hand, the high energy degrees of freedom may manifest themselves during infinitesimal time intervals. The lack of continuity of the evolution operator (51) that results in the fact that the effective $NN$ interaction cannot be parametrized as an instantaneous interaction means that the effects of the high energy degrees of freedom on low energy nucleon dynamics cannot be neglected, and hence one cannot neglect the subspase $H_{in}$ of high momentum states in describing low energy dynamics. Being of the same structure as the space $H_p$, the space $H_{in}$ can be only a subspace of the whole Hilbert space describing high energy states of the underlying theory, while $H_p$ is the subspace of the relevant nucleon states. For describing low energy nucleon dynamics it is sufficient to deal only with amplitudes describing transition between states belonging to $H_p$. However, in processes that are described by these amplitudes the high energy degrees of freedom may come into play. In fact, being infinitesimal from the point of view of low energy nucleon dynamics, the duration time of processes in which high energy degrees of freedom manifest themselves can be finite in the scale of the quark-gluon dynamics. This is the physical cause of the discontinuity of the evolution operator describing low energy nucleon dynamics that results in the fact that low energy nucleon dynamics is non-Hamiltonian.

# OVERVIEW

We have made an analysis of the dynamical situation in the EFT of nuclear forces connected with nonlocality in time of the effective $NN$ interaction. We have shown that this nonlocality gives rise to a lack of continuity of the evolution operator: The evolution operator describing low energy nucleon dynamics satisfies the requirement of the physical continuity, but is not strongly continuous. This discontinuity in turn implies that the Hilbert space of nucleon states cannot be realized as the space $L^2(M)$ (we neglect the spins of the nucleons). This space turns out to be insufficient for describing states of the system. In this case we have to deal with the space $H_p \bigoplus H_{in}$. The subspace $H_p$ can be realized as the space $L^2(M)$, while $H_{in}$ can be realized as the space $L^2(M_\infty)$ constructed by letting the scale $v$ in $L^2(M_v)$ to infinity. Thus $H_{in}$ describes the states corresponding to the infinite energy scale. In a theory with disparate energy scales such states correspond to underlying physics whose energy scale is regarded as infinity. In the case of low energy nucleon dynamics the underlying physics is the quark-gluon dynamics, and $H_{in}$ is the nucleon subspace of the Hilbert space describing this physics. Thus the physical interpretation of the fact, that the Hilbert space we have to use in describing low energy nucleon dynamics should be of the form (61), is obvious: The scales associated with confinement and chiral symmetry are so high that it is natural to regard the states describing the quark and gluon degrees of freedom as unobservable states corresponding to the infinite energy scale, and to formulate the theory of nuclear forces only in terms of the degrees of freedom which emerge after quarks and gluons are confined in bound states. In the case of the EFT of nuclear forces such high energy degrees of freedom as heave vector mesons, $\Lambda's$ and antinucleons are also integrated out, and hence the corresponding states should also be described by the space $H_{in}$.

The above states are not observable in low energy regime and should be regarded as physically nonrealizable. This implies that, if at time $t_0$ the state of a nucleon system was physically realizable, then the probability of finding the system at time $t$ in any state $|\psi\rangle \in H_{in}$ is zero. As we have shown, this really takes place: the evolution operator leaves $H_p$ invariant. Despite this fact, one cannot ignore the subspace $H_{in}$ in describing low energy nucleon dynamics. This has been demonstrated by using our separable model. As we have shown, in the limit $t_2 \to t_1$ the composition law (3) is reduced to Eq. (66) which is written in terms of the operator $\tilde{R}(\theta_2, \theta_1)$. This operator does not leave the space $H_p$ invariant, and in the third nonlinear term of the right-hand side of Eq.(66) we deal only with intermediate states with infinite energies. The state vectors belonging to the space $H_{in}$ also manifest themselves in the generalized dynamical equation (10) in the limit $t_2 \to t_1$. The asymptotic behavior of $\tilde{S}(t_2, t_1)$ in this limit plays a key role in the description of the dynamics of a quantum system: Knowledge of this behavior is sufficient to construct the evolution operator by using the generalized dynamical equation and representation (8). This behavior is determined by the operator $H_{int}(t_2, t_1)$. As we have noted, only the behavior of this operator at infinitesimal duration times $\tau = t_2 - t_1$ of interaction is relevant. In contrast, in the canonical formalism, in order to describe the dynamics of a quantum system one has to start from the processes associated with an instantaneous interaction. It should be emphasized that within the GQD we deal with a new type of nonlocality. In fact, the ordinary way of nonlocalization of a QFT consists in introducing a nonlocal form factors that depend on parameters determining a scale of nonlocality. As for the operator $H_{int}(t_2, t_1)$, only its values in the infinitesimal neighborhood of the point $t_2 = t_1$ are relevant and hence the scale of its nonlocality in time is infinitesimally small, i.e., in this case we deal with some quasilocal operators. This is very important from the point of view of applications to QFT where nonlocalization aimed at resolving the problem of the UV divergences leads to a loss of Lorentz invariance or unitarity. An important feature of the GQD is that it allows one to spread interactions in time, in the above sense, without loosing unitarity. This opens new possibilities for solving the problem of UV divergences in quantum field theory. We have demonstrated this possibility by using the example of the EFT of nuclear forces at leading order. As we have seen, after renormalization low energy nucleon dynamics in this theory is governed by the generalized dynamical equation with the nonlocal-in-time interaction operator (47). Thus the GQD allows one to formulate the EFT of nuclear forces as a perfectly consistent theory free from UV divergences. In contrast with the standard EFT approach, where the two-nucleon $T$ matrix is obtained by summing bubble diagrams and performing regularization and renormalization procedures, in this case the theory leads to the same results being finite at all stages. Note in this connection that, as Dyson [17] pointed out, one may expect that in the future a consistent formulation of quantum field theory will be possible, itself free from infinities, and such that a Hamiltonian formalism may in suitably idealized circumstances be deduced from it. By using the above illustrative example, we have demonstrated that the GQD may be such a formalism, and this gives us the hope that it will a useful tool for solving many problems in quantum field theory.